\documentclass[fleqn,usenatbib]{mnras}

\usepackage{newtxtext,newtxmath}
\usepackage[T1]{fontenc}
\usepackage{ae,aecompl}

\newcommand{\msun}{\mbox{$\,{\rm M}_\odot$}}

\usepackage{graphicx}	
\usepackage{amsmath}	
\usepackage{amssymb}	

\title[Stream disruption]{Quantifying tidal stream disruption in a simulated Milky Way}

\author[E. Sandford et al.]{Emily Sandford,$^{1}$\thanks{E-mail: esandford@astro.columbia.edu}
Andreas H. W. K\"{u}pper,$^{1}$\thanks{Hubble Fellow}
Kathryn V. Johnston$^{1}$,
\newauthor and J\"{u}rg Diemand$^{2}$
\\
$^{1}$Department of Astronomy, Columbia University, 550 West 120th Street, New York, NY 10027, USA\\
$^{2}$Institute for Computational Science, University of Z\"{u}rich, 8057 Z\"{u}rich, Switzerland
}

\date{Accepted 2017 May 19. Received 2017 May 18; in original form 2017 January 30}

\pubyear{2017}

\begin{document}
\label{firstpage}
\pagerange{\pageref{firstpage}--\pageref{lastpage}}
\maketitle

\begin{abstract}

Simulations of tidal streams show that close encounters with dark matter subhalos induce density gaps and distortions in on-sky path along the streams. Accordingly, observing disrupted streams in the Galactic halo would substantiate the hypothesis that dark matter substructure exists there, while in contrast, observing collimated streams with smoothly varying density profiles would place strong upper limits on the number density and mass spectrum of subhalos. Here, we examine several measures of stellar stream ``disruption" and their power to distinguish between halo potentials with and without substructure and with different global shapes. We create and evolve a population of 1280 streams on a range of orbits in the Via Lactea II simulation of a Milky Way-like halo, replete with a full mass range of $\Lambda$CDM subhalos, and compare it to two control stream populations evolved in smooth spherical and smooth triaxial potentials, respectively. We find that the number of gaps observed in a stellar stream is a poor indicator of the halo potential, but that (i) the thinness of the stream on-sky, (ii) the symmetry of the leading and trailing tails, and (iii) the deviation of the tails from a low-order polynomial path on-sky (\textit{``path regularity"}) distinguish between the three potentials more effectively. We furthermore find that globular cluster streams on low-eccentricity orbits far from the galactic center (apocentric radius $\sim 30-80$ kpc) are most powerful in distinguishing between the three potentials. If they exist, such streams will shortly be discoverable and mapped in high dimensions with near-future photometric and spectroscopic surveys.

\end{abstract}

\begin{keywords}
Galaxy: halo -- galaxies: haloes -- galaxies: kinematics and dynamics -- general: dark matter.
\end{keywords}

\section{Introduction} \label{sec:intro}

According to the prevailing $\Lambda$CDM cosmological model, structure formation in the Universe proceeds hierarchically: early in the Universe's lifetime, dark matter collapses on small scales to form small clumps, and over time, these small clumps merge to form larger and larger structures \citep{white78}. Numerical simulations of this process predict that evidence of this hierarchical history should abound in the present-day Universe, because the cores of merged small dark matter clumps survive as ``subhalos" within the larger dark matter halos of galaxies (see e.g. \citealt{diemand08,springel08}). 



However, observations of the Milky Way's halo challenge these predictions. To date, even with the advent of deep imaging surveys such as SDSS and the Dark Energy Survey, only a few tens of satellite galaxies have been detected around the Milky Way  (see e.g. \citealt{des15,bechtol15,koposov15,laevens15a,laevens15b,martin15,kimjerjen15a,kimjerjen15b,belokurov14}; see \citealt{mcconnachie12} for a review of earlier discoveries). Dark matter-only simulations predict that the Milky Way should host at least an order of magnitude more subhalos than are accounted for by these visible satellites \citep{moore99,klypin99}. There are several possible explanations for this discrepancy: perhaps the Milky Way is a statistical outlier among galaxies of its kind \citep{geha14}, or $\Lambda$CDM is inadequate to describe the universe at small scales (see e.g. \citealt{bode01, hui16}). A third possibility is that numerous low-mass dark subhalos do exist in the galaxy but fail to form stars due to baryonic processes that are not included in dark matter-only simulations of structure formation (see e.g. \citealt{quinn96, brooks13}).

Subhalos which cannot be directly observed may nonetheless be detected by their gravitational effects. Subhalos around strong gravitational lens galaxies, for example, distort lensed images and cause anomalous flux ratios in multiply-imaged sources (see e.g. \citealt{mao98,vegetti09a}). Subhalos as small as $10^7\msun$ may be detectable around lens galaxies within the next few years with large observational surveys to discover new lenses, such as LSST, and new high-resolution and high-sensitivity instruments, such as ALMA and SKA  \citep{hezaveh13, vegetti09b}.

Within the Milky Way, dark subhalos are most promisingly detectable by their gravitational effects on stellar streams (see e.g. \citealt{ibata02, johnston02}). Such streams form when satellite galaxies or globular clusters are tidally disrupted and stretched into long, thin streams. Streams from globular clusters, in particular, are useful probes of the Galactic potential because of their low velocity dispersion and consequent sensitivity to small gravitational perturbations. Many numerical studies have demonstrated that close encounters with dark matter subhalos leave observational traces on stellar streams, most notably dynamical heating (e.g. \citealt{siegalgaskins08,carlberg09}) and gaps in surface density (e.g. \citealt{yoon11,carlberg11,carlbergetal12,ngan15,erkal15,sanders16}).


However, more recent work has hinted that disrupted stellar streams in the Galaxy may not constitute straightforward proof of low-mass dark subhalos. Firstly, streams may be disrupted even in the absence of dark matter substructure. \cite{pearson15}, for example, find that halo triaxiality can induce ``fanned" morphologies, in which stream particles disperse in two dimensions as they move away from their globular cluster progenitor. \cite{ngan16} compare streams evolved in smooth and lumpy spherical and triaxial potentials and find that introducing asymmetry in the overall halo shape causes streams to be more widely dispersed than adding subhalos does. \cite{apw16b} model the Ophiuchus stream in a potential with a rotating bar component and find that the bar induces a large dispersion in both stellar position and velocity in models. \cite{pearson17} model the Palomar 5 stream with a rotating bar and demonstrate that the bar can create density gaps along the stream; \cite{erkal16arxiv} find that the bar induces asymmetric density variations to their modeled Palomar 5 stream. \cite{amorisco16} find that giant molecular clouds can create gaps and clumps in N-body streams during close flybys, just as dark matter subhalos do. Finally, \cite{kupper08} demonstrate that periodically spaced ``epicyclic overdensities" arise along tidal streams even when the tidal field is constant, as when the stream progenitor is on a circular orbit, in the absence of halo substructure. 

Secondly, the observational evidence for subhalo-induced disruption of Galactic streams has proven stubbornly difficult to interpret. For example, there have been several conflicting analyses of observations of the Palomar 5 stream, particularly with regard to whether it has significant under- or overdensities. \cite{carlbergetal12}, using a gap-finding wavelet analysis on SDSS photometric data, identify at least five significant gaps along the stream; \cite{kupper15} find twenty-four overdensities in the same data using an independent statistical method. In newer CFHT data, \cite{ibata16} use the same wavelet method as \cite{carlbergetal12} to identify only one significant overdensity and no significant gaps. \cite{erkal16arxiv}, in analyzing the same CFHT data, adopt a non-parametric modeling approach and detect two significant gaps, and two significant peaks, in the Pal 5 stream's density profile. Meanwhile, \cite{bovy16}, instead of counting individual gaps or peaks at discrete scales, analyze the power spectrum of perturbations in density along the Pal 5 stream and conclude that the power on large scales is consistent with that expected from a $\Lambda$CDM subhalo population, while the power on scales $\lesssim 5^{\circ}$ is obscured by noise.

The number of conflicting conclusions as to the reality of subhalo-induced gaps in Galactic streams suggests that perhaps the observational data are not yet good enough to decide the question one way or the other. In any case, they emphasize the need to carefully account for the large observational uncertainties in the data: \cite{thomas16}, for example, note that even a perfectly smooth artificial stream can be found to have significant overdensities when observed against a realistic SDSS-like background. The conflicting results of gap searches also suggest that gaps are not the ``smoking gun" of dark matter substructure interaction that we had once hoped and that we should search for other signals of subhalo influence on streams. 

In this work, we investigate such signals, in hopes of identifying stream features which are more unambiguously attributable to subhalos. We present the Via Lactea Cauda stream sample, which consists of 1280 stellar streams evolved from Palomar 5-like globular cluster progenitors on a range of orbits in the full Via Lactea II (VL2) simulation of a Milky Way-like halo \citep{diemand08}. This halo is time-evolving, triaxial, and populated with a full mass range of $\Lambda$CDM subhalos. We compare these Cauda streams to control streams evolved from identical initial conditions in two smooth potentials (triaxial and spherical) fit to VL2. 


Of this simulated data set, we ask the following specific questions:

\begin{enumerate}
    \item How ``disrupted" are the Cauda streams relative to their smooth potential counterparts?
    \item Which measures of stream disruption are most useful for distinguishing between the VL2 and smooth potentials?
    \item Does a stream's level of disruption correlate with its position in the potential? If so, which streams are best placed to distinguish between the VL2 and smooth potentials?
\end{enumerate}

In \S \ref{sec:data}, we present details of the VL2 simulation and the generation of the Cauda and control streams. In \S \ref{sec:analysis}, we define five measures of stream ``disruption." In \S \ref{sec:results}, we evaluate the degree of stream disruption in each potential by each of the five measures, with particular focus on streams on Palomar 5-like orbits, and find that the most promising measures for distinguishing between LCDM and smooth potentials are on-sky path regularity, stream symmetry, and stream thinness on-sky. We also find that, when a realistic level of background noise is introduced, streams on circular orbits farther out in the potential than Palomar 5 ($r_{apo} \sim 30-80$ kpc) are best placed to distinguish between the different potentials. We emphasize that Palomar 5 is less useful than such streams even in our simulations, in which there is no disk to destroy close-in subhalos \citep{donghia10}. In \S \ref{sec:discussion}, we interpret our results in the context of the observational and theoretical work referenced above and consider the prospects of observing the telltale signs of subhalo interactions presented here. We summarize in \S \ref{sec:conclusions}.

\section{Data} \label{sec:data}

\subsection{The Via Lactea II Simulation}

The streams presented in this paper are grown and evolved in a modified recomputation of the Via Lactea II (VL2) simulation \citep{diemand08}.  VL2 is a dark-matter-only simulation that follows the formation of a Milky Way-like halo from $z = 104.3$ to $z = 0$ in a $\Lambda$CDM universe with cosmological parameters drawn from \textsc{WMAP} data. In this work, we recompute the last 6\,Gyr of evolution, starting from a snapshot of the original VL2 at $z = 0.70$, using the same \textsc{PKDGRAV2} tree code as the original simulation. In this 6\,Gyr, the VL2 main halo experiences no major mergers but ongoing minor mergers.

The computations were carried out on SuperMUC at LRZ Garching, Germany, and on the zBox supercomputer at the University of Zurich, Switzerland. We use essentially the same resolution as the original VL2, with 1.1 billion particles of $4100\msun$ each and a force resolution of $40$~pc. Unlike in the original VL2, we fix the time steps of the simulation to 1\,Myr for computational convenience, which only affects the accuracy of a few particles in the very center of the main halo. 

At $z=0$, the main halo has a mass of $M_{200} = 1.9\times10^{12}\,\msun$ within $R_{200} = 402$~kpc (where $R_{200}$ is defined as the radius enclsing a mass density of 200 times the cosmic average dark matter density). Hence, it appears to be a bit more massive than current estimates of the Milky Way's mass (see e.g. \citealt{bhattacharjee14,gibbons14,deason12,gnedin10}). The mass within the inner 150\,kpc of the $z=0$ VL2 halo, the region of interest in the present investigation, is $M_{tot}(<150\,\mbox{kpc}) = 1.16 \times 10^{12}\msun$ \citep{bonaca14}.


\subsection{Via Lactea Cauda} \label{subsec:VLC}

Into this VL2 re-simulation, we inject 1280 particles that act as our test globular clusters. These test globular clusters, or ``progenitors,'' each release 1 stream particle at each Lagrange point at each time step to grow a leading and a trailing tail. By the end of the simulation, each progenitor particle has released $2 \times 6000 = 12,000$ stream particles, leaving us with 15 million stream particles for analysis.

Each globular cluster particle has a constant mass of $20,000\msun$ (modeled after the Palomar 5 globular cluster), whereas the stream particles, which represent stars in the resulting stream (typical mass $\sim 0.4\msun$), are assumed to be massless. The gravitational potential of each globular cluster particle is modeled as a smoothed point mass potential with a softening length of 20 pc ($F_{\textrm{grav}} = \frac{GM}{R^2 + c^2} m_{\mathrm{test}}$, where $M$ is equal to the cluster mass, $20,000\msun$, and $c$ is equal to 20 pc). In this mass scheme, we effectively neglect star-star interactions within the streams, but accurately simulate the orbital motion of globular clusters and stream stars in the gravitational potential of the VL2 dark matter particles and the globular cluster particles. Most importantly, we follow the effects of encounters between streams and dark matter subhalos. At $z=0$, the simulation resolves more than $40,000$ such subhalos within $R_{200}$, ranging in mass from $10^6\msun$ to $10^9\msun$ with roughly equal mass per mass decade. We and others have found that the abundance and radial distribution of subhalos in VL2 is accurate to within $\sim10\%$ down to a subhalo mass below $10^6\msun$, corresponding to $\sim 200$ dark matter particles (see e.g. \citealt{diemand07}). While this is insufficient to resolve the internal properties of a subhalo, it is sufficient to model its external effects on streams except in the case of direct subhalo impacts; however, such direct encounters (with an impact parameter less than or equal to the scale radius of the subhalo) comprise only $7\%$ of the encounters tracked in the simulation.

We generate the massless stream particles using a variant of the commonly applied ``streakline" or ``particle spray'' method \citep{varghese11, kupper12, lane12, bonaca14, amorisco14, gibbons14, fardal14}. This method assumes that stars evaporate from star clusters due to two-body relaxation and mild tidal shocks \citep{baumgardt03, kupper08, kupper10}. Instead of using time consuming $N$-body integrations for the dynamics of stars within the cluster, this approach creates the ``stars'' (or stream tracer particles) in the moment of their escape from the cluster. This reduces the computational costs dramatically, since only the gravitational influence of the potential of the host galaxy and of the cluster as a whole must be taken into account in calculating the motions of the stream particles.

Since stars primarily escape from star clusters through the Lagrange points of the respective cluster-host galaxy system \citep{king62}, the stream particles are initialized in these locations with small offset velocities to avoid re-accretion by the cluster. The Lagrange points are approximately located along the connection line between the cluster center and host galaxy center at a distance of one tidal radius from the cluster center. The tidal radius can be calculated as
\begin{equation}
r_t = \pm\left(\frac{GM}{\Omega^2-\partial^2\Phi/\partial R^2}\right)^{1/3},
\end{equation}
where $M$ is the mass of the cluster, $\Omega$ is its angular velocity with respect to the host galaxy center, and $R$ is its galactocentric distance. Note that $r_t$ depends on the second partial derivative of the host galaxy potential, $\Phi$, with respect to $R$. Since this quantity is not known beforehand for a live dark matter halo in formation, we approximate this quantity with an average value derived from the original VL2 simulation. For this purpose, we fit a third-order polynomial to the spherically averaged $\partial^2\Phi/\partial R^2$ profile of the $z=0$ and the $z=0.70$ snapshots of the VL2 simulation, binned in 100 linearly spaced bins out to 200 kpc from the center of the main halo. Since the main halo does not experience any major mergers within the simulation time, this force gradient is remarkably stable over time: the difference between the snapshots is much smaller than 1\% for $R>10$\,kpc, and is of the order of 10\% for $R<10$\,kpc. Since $r_t$ has a only a weak dependence on $\partial^2\Phi/\partial R^2$, we use the average between the two snapshots to estimate its value for each cluster particle at each time step.

To generate a realistic level of dispersion among the stream particles, we add a spatial offset drawn from a normal distribution with a standard deviation of $0.25 r_t$, and a velocity offset drawn from a Maxwell-Boltzmann distribution with a dispersion equal to the velocity dispersion $\sigma = 2$\,km\,s$^{-1}$ of the progenitor (chosen to match the velocity dispersion of Palomar 5), to each stream particle as it is initialized.\footnote{In analyzing the Cauda streams, we discovered that, in our VL2 re-simulation, the integration of the progenitor particle orbits with PKDGRAV2 was done in a comoving reference frame, while the offset velocities of the tracer particles released from the progenitor particle at every time step were calculated in a non-comoving frame. We compensate for this discrepancy when generating the smooth potential comparison streams described in \S \ref{subsec:smooth_potential} by multiplying the tracer particle offset velocities by the square of the scale factor. While we could not re-generate the VL2 streams to correct for the discrepancy from the outset, we ran tests on smooth potential streams, varying the stream particle offset velocities within the range of the above multiplication by the square of the scale factor, and confirmed that the qualitative nature of the streams did not change. In particular, the average number of detected gaps per stream by the procedure described in \S \ref{subsubsec:gaps} remained the same.}

\begin{figure}
\centering
\includegraphics[width=0.45\textwidth]{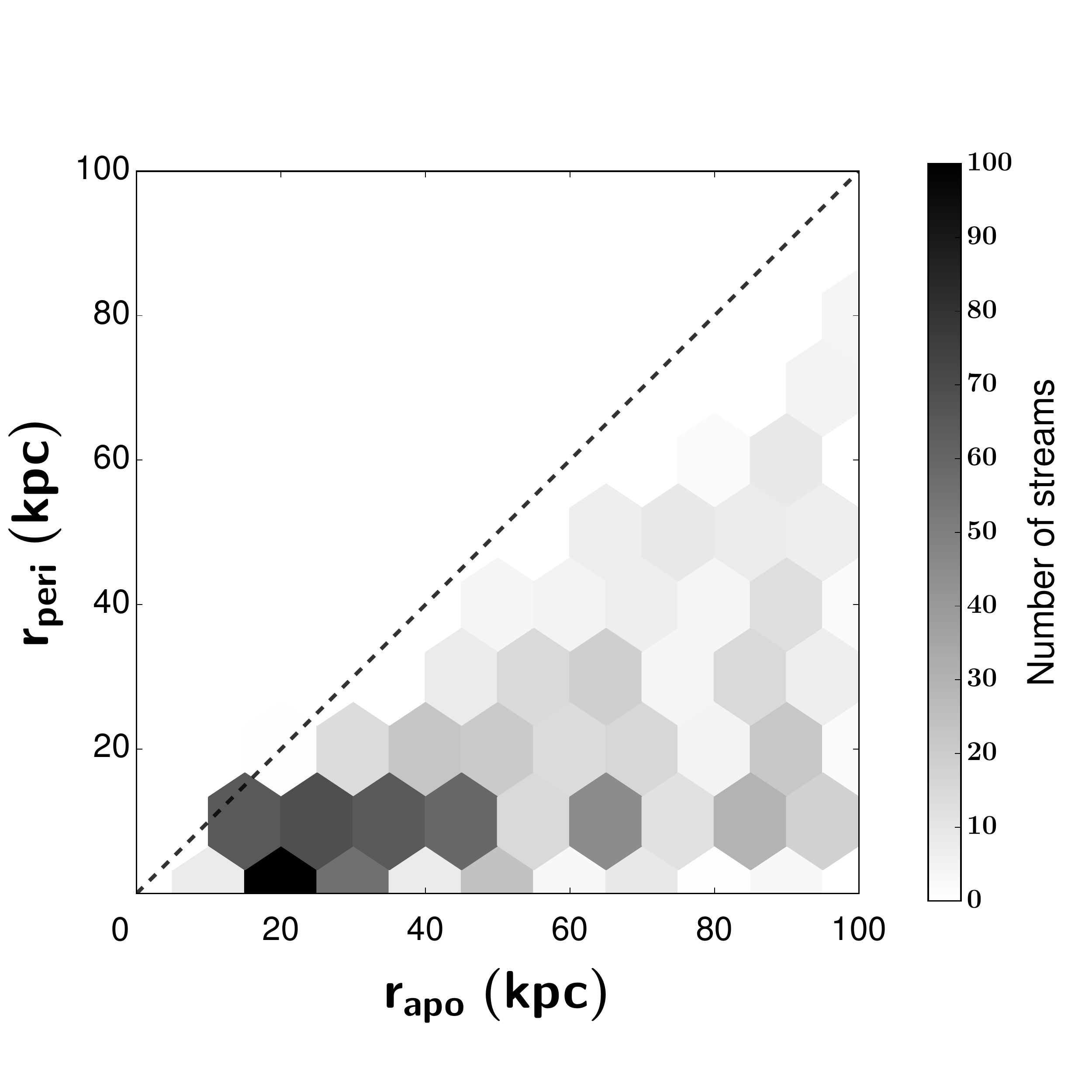}
\caption{The distribution of orbits of the globular cluster progenitors of the Via Lactea Cauda streams.}
\label{fig:spatialbins}
\end{figure}

\begin{figure}
\centering
\includegraphics[width=0.45\textwidth]{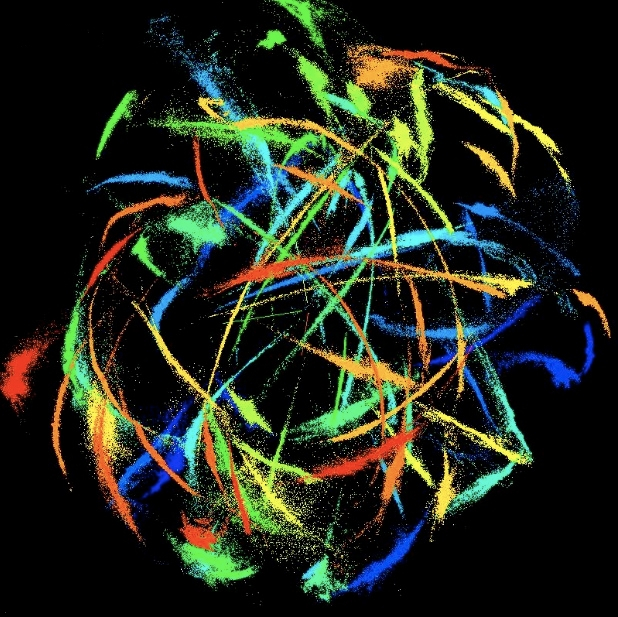}
\caption{A snapshot of a subset of the Via Lactea Cauda streams during their evolution. This view is 130 kpc by 130 kpc.}
\label{fig:snapshot}
\end{figure}

The globular cluster particles are initialized at their orbital apocenters, which range from 15 to 150 kpc, and are given initial velocities ranging from 0.25 to 1 times the local circular velocity, $V_C = \sqrt{r \frac{\mathrm{d}\Phi}{\mathrm{d}r}}$, at $z=0.70$, resulting in a range of eccentricities. The orbital distribution of the globular cluster progenitors is shown in Fig.~\ref{fig:spatialbins}. A typical snapshot of a subset of the Via Lactea Cauda streams can be seen in Fig.~\ref{fig:snapshot}. 

\subsection{Smooth potential control streams} \label{subsec:smooth_potential}
To isolate the effects of close encounters with dark matter subhalos on the present-day Via Lactea Cauda streams, we generate two analogous control stream populations in smooth, time-evolving potentials fit to the Via Lactea II halo. Following the fitting procedure of \cite{bonaca14}, we fit triaxial and spherical logarithmic potentials of the form:

\begin{equation}\label{eq:logpotential}
\Phi_{log} = V_c^2 \ln(r^2 + R_c^2)
\end{equation}

where $V_c$ is the circular velocity and $R_c$ is the core radius. The radial coordinate $r$ is defined to allow for halo triaxiality:

\begin{equation}
r^2 = C_1 x^2 + C_2 y^2 + C_3 xy + \left( \frac{z}{q_z} \right)^2
\end{equation}

The constants $C_1$, $C_2$, and $C_3$ are defined in \cite{law10}:

\begin{equation}
\begin{split}
C_1 = \left( \frac{\cos{\phi}}{q_1} \right)^2 + \sin^2{\phi}
\\
C_2 = \cos^2{\phi} + \left( \frac{\sin{\phi}}{q_1} \right)^2
\\
C_3 = 2\sin{\phi}\cos{\phi} \left( \frac{1}{q_1^2} - 1 \right)
\end{split}
\end{equation}

The parameters fit to the VL2 halo potential are $V_c$, $R_c$, the dimensionless ratio of the x- and y-axes $q_1$, the dimensionless ratio of the z- and y-axes $q_z$, and the rotation angle around the z-axis $\phi$. In the case of the triaxial potential fit, $q_1$ and $q_z$ are allowed to vary, while in the spherical potential fit, they are fixed to 1, such that $r^2$ reduces to $x^2 + y^2 + z^2$. \cite{bonaca14} found that a triaxial potential of the form of equation~\ref{eq:logpotential} approximates the true VL2 potential quite well, with residuals on the order of 3\%. 

We account for time evolution over the final 6 Gyr of the VL2 halo potential evolution by fitting these parameters to ten snapshots of VL2 equally spaced in time over these 6 Gyr, then approximating the evolution of each parameter with a linear fit to the data. The best-fit parameters and their dependence on time are given in Table~\ref{table:pot_params}. In general, the potential parameters undergo very slow evolution; none changes by more than 10\% over the course of the 6 Gyr.

To generate streams in these smooth potentials, we adopt the present-day ($t = 0,\ z = 0$) positions and velocities of the 1280 globular cluster particles from the Via Lactea Cauda simulation. We integrate these globular cluster particles backward along their orbits for 6 Gyr in the triaxial and spherical potentials. We then integrate them forward to $t=0$ again, generating streakline streams along the way by the procedure described in section~\ref{subsec:VLC}. This results in two control populations of 1280 streams, one each from the triaxial and spherical smooth potentials, evolved to match the final positions and velocities of the Via Lactea Cauda streams but not subject to any disruption by orbiting subhalos. 

\begin{table*}
\centering
\caption{Parameters of the triaxial and spherical logarithmic potential fits to VL2. $t$ ranges from -6000 Myr, the starting point of the Via Lactea Cauda simulation, to 0 Myr, the present day.}
\label{table:pot_params}
\begin{tabular}{lll}
Parameter & Triaxial potential & Spherical potential  \\
\hline
$V_c$ [km/s]  &  $140.59867 + (3.6\mathrm{e}{-4}\ \mathrm{Myr}^{-1}\ t)$ & $140.42744 + (3.5\mathrm{e}{-4}\ \mathrm{Myr}^{-1}\ t)$   \\
$R_c$ [pc]  &  $5425.791 + (5.5\mathrm{e}{-2}\ \mathrm{Myr}^{-1}\ t)$ & $6539.707 + (6.6\mathrm{e}{-2}\ \mathrm{Myr}^{-1}\ t)$  \\
$q_1$  &  $1.155496 - (1.0\mathrm{e}{-6}\ \mathrm{Myr}^{-1}\ t)$  & 1 \\
$q_z$  &  $1.326417 + (2.0\mathrm{e}{-6}\ \mathrm{Myr}^{-1}\ t)$  & 1 \\
$\phi$ [rad] &  $1.434162 - (2.0\mathrm{e}{-6}\ \mathrm{Myr}^{-1}\ t)$ & -
\end{tabular}
\end{table*}

\subsection{Stream density profiles}\label{subsec:densityprof}
To facilitate the analysis of stream disruption described below in~\ref{subsec:measures}, we measure the linear density along each of the VL2 and control smooth potential streams. First, we convert the present-day Cartesian space coordinates  $(x,y,z)$ of the stream particles to Galactic coordinates $(l,b)$, the coordinates which are observed for real Galactic streams. 

We then rotate the Galactic coordinates $(l, b)$ of each stream such that the globular cluster particle progenitor lies at the pole (see figure~\ref{fig:methods_gaps_smooth}, first panel). We term these rotated longitude and latitude coordinates $\lambda$ and $\beta$, respectively. After this rotation, it is clear that the rotated latitude coordinate $\beta$ measures angular distance along the stream; we assign positive $\beta$ to correspond to the leading tail and negative $\beta$ to correspond to the trailing tail. We calculate a one-dimensional kernel density estimate (KDE) along the rotated latitude coordinate $\beta$ at evenly spaced intervals of $0.01^{\circ}$, with a Gaussian kernel of size $0.1^{\circ}$ (chosen to correspond to the minimum scale over which we search for stream gaps; see~\ref{subsubsec:gaps}). This procedure smooths the discrete particle positions into a continuous 1D density profile along the length of the stream. Example linear density profiles are shown in the upper right panels of figures~\ref{fig:methods_gaps_smooth} and~\ref{fig:methods_gaps_gappy}.

\section{Analysis Methods} \label{sec:analysis}

\subsection{Quantifying stream disruption} \label{subsec:measures}
To evaluate the comparative level of stream disruption in each potential for a data set this large, we must choose objective, quantitative measures of ``disruption" which can be implemented algorithmically and evaluated on the VL2, smooth triaxial, and smooth spherical potentials. Broadly, we expect ``intact" streams to be long, thin, smooth, and symmetrical, with simple, regular paths on-sky; we expect the opposite for streams disrupted by subhalo encounters. 

Previous work has quantified disruption in terms of the cumulative dynamical heating of the stream resulting from close subhalo encounters, as well as the number of gaps left in the stream by individual subhalo flybys. We adopt these measures as well; we parametrize the degree of dynamical heating in terms of the ``thinness" of the stream on-sky (see~\ref{subsubsec:thinness}, below), and we count the statistically significant gaps in linear density along the stream (see~\ref{subsubsec:gaps}). 

Additionally, however, we consider three other measures of stream disruption: the angular length of the stream on-sky (\ref{subsubsec:length}), the symmetry of the leading and trailing tails (\ref{subsubsec:symmetry}), and the on-sky \textit{``path regularity"} of the leading and trailing tails (\ref{subsubsec:straightness}). Descriptions of our five measures and our procedure for evaluating them follow. In~\ref{subsec:discussion_measures}, we discuss our chosen measures in the context of the contrasting power-spectrum approach taken by \cite{bovy16}.

Finally, in~\ref{subsec:noise}, we add background stars to our simulated streams and reevaluate the detectable level of stream disruption if the streams were observed at a comparable signal-to-noise ratio to the recent Palomar 5 stream observations of \cite{ibata16}.

\subsubsection{Length} \label{subsubsec:length}
\begin{figure*}
\begin{center}
\includegraphics[width=\textwidth]{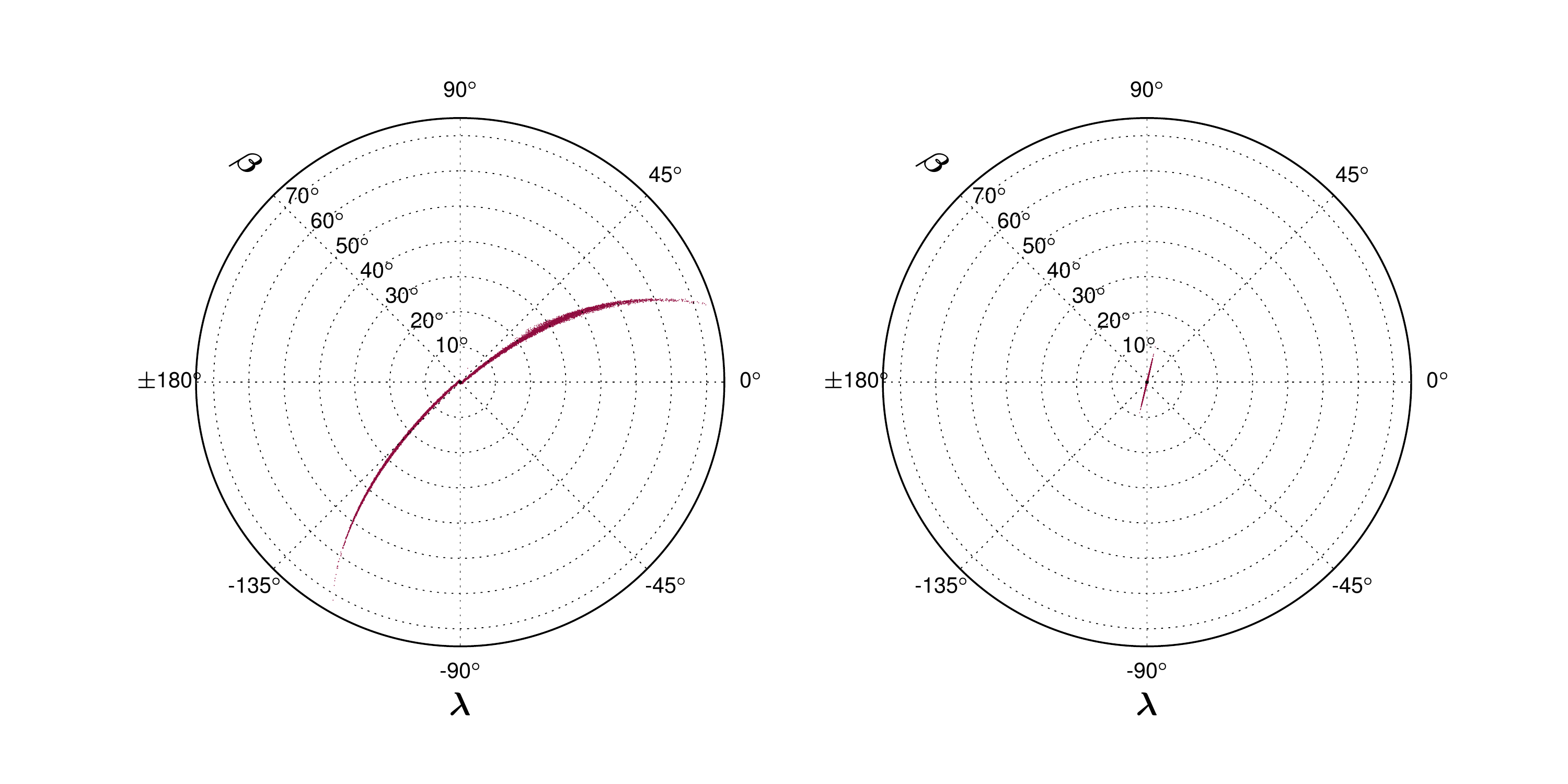}
\caption{Two streams evolved in the spherical smooth potential. Left: a stream measuring $127.5^{\circ}$ on-sky; right: a stream measuring $19.2^{\circ}$ on-sky. Both streams are on low-eccentricity ($e < 0.25$) orbits, but the short stream has an apocentric radius roughly 10 times that of the long stream and has completed fewer orbits in 6 Gyr of evolution. Stream length is strongly correlated with apocentric distance, but not with subhalo interactions, as we show in~\ref{sec:results}.}
\label{fig:methods_length}
\end{center}
\end{figure*}

To measure the length of an individual stream, we use a pairwise distance routine to calculate the maximum on-sky angular separation between any two particles in the stream. Fig.~\ref{fig:methods_length} illustrates a long and a short stream.

\subsubsection{Thinness} \label{subsubsec:thinness}
To judge stream thinness on-sky, we apply principal component analysis (PCA) to the rotated coordinate positions $(\lambda,\beta)$ of the stream particles. PCA attempts to project the two-dimensional structure of the stream into a lower-dimensional space: if all the stream particles lay perfectly on a line, for example, then only one coordinate would be necessary to describe the stream, and PCA would return the stream particle positions in that coordinate (i.e., the distance of each stream particle from the origin along the line). More generally, PCA projects the data onto a number of orthogonal ``components" less than or equal to the original dimensionality of the data points (in our case, two), choosing the first component to capture the maximum possible variance across the data set, and then subsequent components to capture the remaining maximum possible variance while remaining orthogonal to the existing components.

\begin{figure*}
\begin{center}
\includegraphics[width=\textwidth]{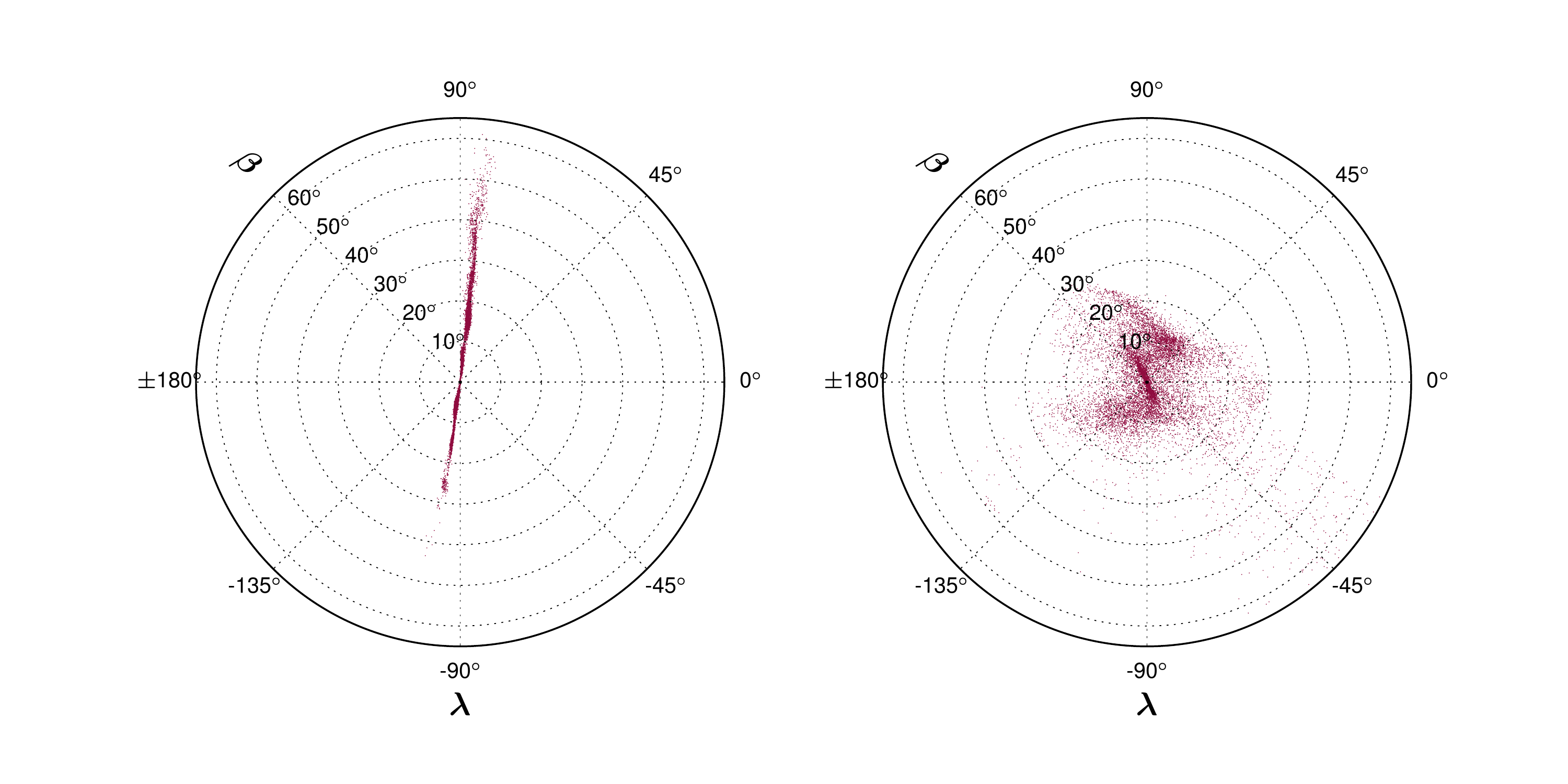}
\caption{Two streams evolved in the VL2 potential. Left: a ``thin" stream whose first principal component explains $99.9\%$ of the variance in the on-sky positions of its constituent particles, leaving only $0.1\%$ of the variance to be explained by the second principal component. The ratio of explained variances in this case is $> 10^3$. Right: a ``non-thin" stream whose first principal component explains only $67.7\%$ of the variance in the on-sky positions of its constituent particles, leaving $32.3\%$ of the variance to be explained by the second principal component. The ratio of explained variances for this stream is roughly 2.}
\label{fig:methods_thinness}
\end{center}
\end{figure*}

None of our streams is a perfect line, so each requires two principal components to fully specify the stream particle positions. To quantify the stream ``thinness," we evaluate the ratio of the variance explained by the first principal component to the variance explained by the second. A ratio $>> 1$ indicates that the stream is much more dispersed along the first component than the second, so it is elongated and ``thin;" a ratio $\simeq 1$ indicates that the stream is roughly equally dispersed in two dimensions, so it is blob-like. Figure~\ref{fig:methods_thinness} illustrates each type of stream. We use the scikit-learn implementation of PCA \citep{sklearn} to choose the orthogonal components and calculate this variance ratio. 

\subsubsection{Thin fraction}
While the above measures (length and thinness) can be evaluated regardless of stream morphology, our other measures of disruption (gaps, symmetry, and path regularity) are only appropriately applied to ``thin" streams. An observer of Galactic streams would only attempt to identify gaps in surface density, for example, along a structure which had already been identified as a tidal stream. Similarly, measuring the leading vs. trailing tail symmetry or path regularity of a bloblike structure is nonsensical. We therefore impose a ``thinness" cutoff and restrict the analyses described in~\ref{subsubsec:gaps} through \ref{subsubsec:straightness} to streams which pass the cutoff. As a cutoff, we conservatively choose an explained variance ratio of 10, motivated by the 50:1 measured length-to-width ratio of the Ophiuchus stream \citep{sesar15}. 

\subsubsection{Number of gaps} \label{subsubsec:gaps}
To count the number of significant gaps per stream, we adopt a wavelet convolution procedure similar to that of \cite{carlbergetal12} and \cite{ibata16} in order to identify places where stream stars have vacated a gap in the stream and piled up on either side. Such a gap appears in the stream density profile as an underdensity bounded on both sides by overdensities and can be identified by convolution of the stream density profile with a gap filter of similar shape (we adopt the \citealt{carlbergetal12} $w_2$ filter).

\begin{figure*}
\begin{center}
\includegraphics[width=\textwidth]{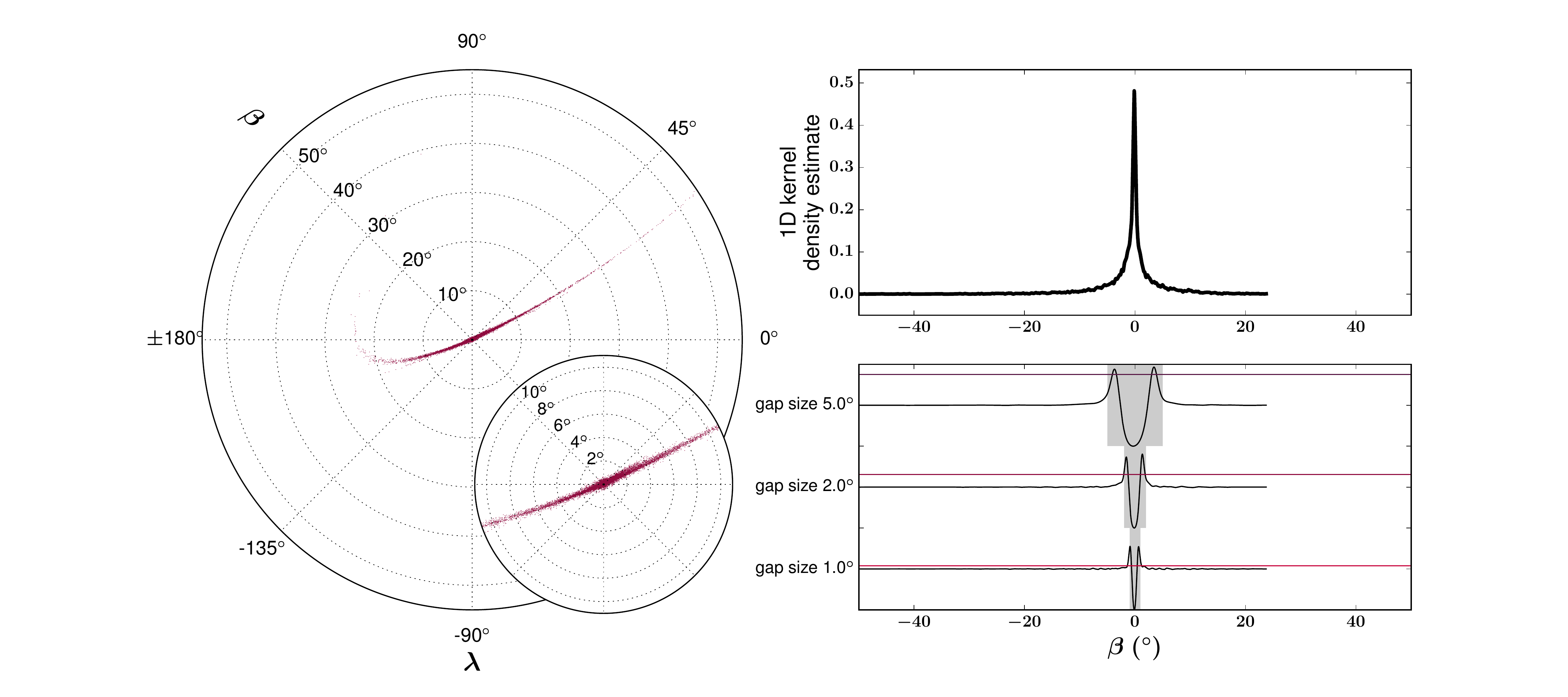}
\caption{Left: A stream evolved in the spherical smooth potential along which no significant gaps are detected. This stream is visibly smooth in its central regions (inset); the apparent gaps at the far ends of the tails, where there are very few stream particles, are insignificant, the effect of low-particle-number noise. Upper right: The linear density profile of this stream, plotted against $\beta$. Lower right: The convolution of the linear density profile with the \protect\cite{carlbergetal12} $w_2$ filter (black curves), scaled to find, from bottom to top, gaps of characteristic width $1^{\circ}$, $2^{\circ}$, and $5^{\circ}$. The colored horizontal lines mark the significance thresholds for each gap scale as determined by the Monte Carlo procedure outlined in~\ref{subsubsec:gaps}. The gray shaded regions were not searched for gaps based on their proximity to the linear density profile peak at the globular cluster position ($\beta = 0$); the convolution evidently has spurious peaks (similar in shape to the $w_2$ filter itself) in these regions due to the density peak. No significant gaps are detected for this stream. The relatively extreme height of this stream's KDE peak (upper right panel) compared to that of figure~\ref{fig:methods_gaps_gappy} is due to this stream's orbital phase; it is at apogalacticon, so it is compressed \citep{kupper10}.}
\label{fig:methods_gaps_smooth}
\end{center}
\end{figure*}

\begin{figure*}
\begin{center}
\includegraphics[width=\textwidth]{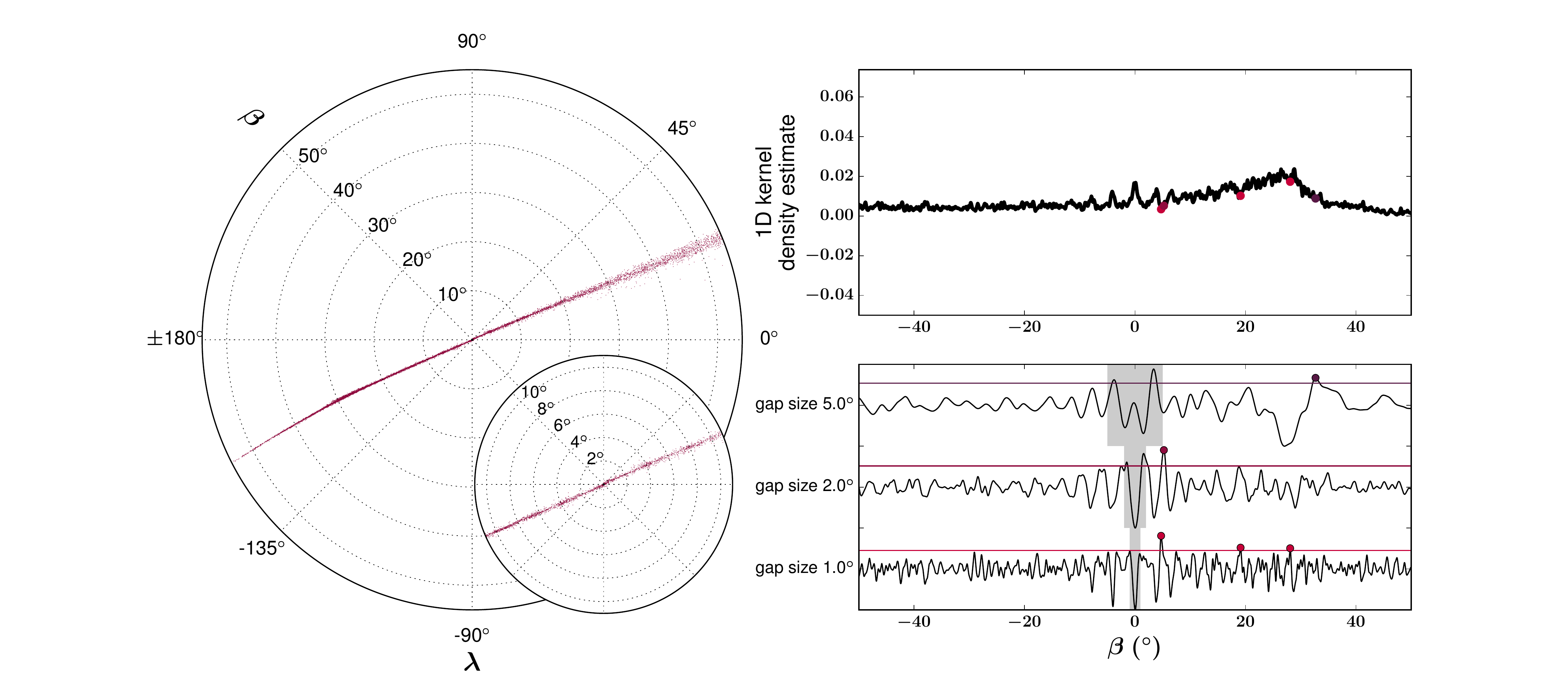}
\caption{Left: Another stream evolved in the spherical smooth potential. Inset: A zoomed-in view of the innermost $20^{\circ}$ of this stream; note the obvious density variations, which appear to be epicyclic over- and underdensities \citep{kupper12} based on their symmetry and regular spacing. Upper right: The linear density profile of this stream, plotted against $\beta$. Lower right: The convolution of the linear density profile with the \protect\cite{carlbergetal12} $w_2$ filter (black curves), scaled to find, from bottom to top, gaps of characteristic width $1^{\circ}$, $2^{\circ}$, and $5^{\circ}$. The colored horizontal lines mark the significance thresholds as determined by the Monte Carlo procedure outlined in~\ref{subsubsec:gaps}. The gray shaded regions were not searched for gaps based on their proximity to the linear density profile peak at the globular cluster position ($\beta = 0$). Colored dots mark the positions of significant peaks identified in the convolution and the significant gaps in the linear density profile to which they correspond. We note that the overdensity at $\beta \simeq 25^{\circ}$ is due to the leading tail of the stream bunching up as it approaches its apocenter.}
\label{fig:methods_gaps_gappy}
\end{center}
\end{figure*}

We measure the linear density profiles of the thin streams by the procedure described in~\ref{subsec:densityprof}. There is a clear qualitative difference between the linear density profiles of streams with visually obvious gaps in surface density (see fig.~\ref{fig:methods_gaps_gappy}, upper right panel) and streams that appear smooth on-sky (fig.~\ref{fig:methods_gaps_smooth}, upper right panel). To quantify the level of density variation along each stream, we follow the procedure of \cite{ibata16}, searching for gaps by convolving the linear density profile with the gap-shaped \cite{carlbergetal12} $w_2$ filter. We scale the filter to search for both sub-degree characteristic widths ($0.1^{\circ}, 0.2^{\circ}, 0.5^{\circ}$) and super-degree characteristic widths ($1.0^{\circ}, 2.0^{\circ}, 5.0^{\circ}$). Peaks in the resulting convolution represent places where the shape of the stream density profile resembles the shape of the $w_2$ filter.

To pick out the statistically significant peaks in the convolution, corresponding to the statistically significant gaps, we define significance thresholds using Monte Carlo analysis, again following \cite{ibata16}. For each stream, we create 100 random ``realizations" of the linear density profile, where, at each $0.01^{\circ}$ interval in $\beta$, we draw from a Gaussian density distribution describing the density at the center of that interval. The mean of this Gaussian is given by the number of particles in the interval, calculated as the value of the one-dimensional KDE times the interval width ($0.01^{\circ}$) times the total number of particles in the stream. The standard deviation is given by the square root of the number of particles in the interval.

We convolve each of the 100 density profile realizations generated by this procedure with the $w_2$ filter at all six gap scales of interest and search for the maximum height of the resulting peaks in the convolution. We take the threshold for gap ``significance" to be the 99th percentile of the maximum peak height over the 100 random realizations. This procedure results in unique significance thresholds for all 1280 streams in each potential at all six gap scales. Any peak in the convolution between the \textit{real} linear density profile of the stream and the $w_2$ filter which exceeded the appropriate significance threshold is counted as ``significant" (see figs.~\ref{fig:methods_gaps_smooth} and~\ref{fig:methods_gaps_gappy}, lower right panels). We exclude peaks within 1 gap scale of the edge of the stream to avoid edge effects and peaks within 1 gap scale of the globular cluster particle position ($\beta = 0$), which we frequently found to be spurious signals, artifacts of the convolution between the filter and the density peak around the globular cluster.

\subsubsection{Symmetry} \label{subsubsec:symmetry}
\begin{figure*}
\begin{center}
\includegraphics[width=\textwidth]{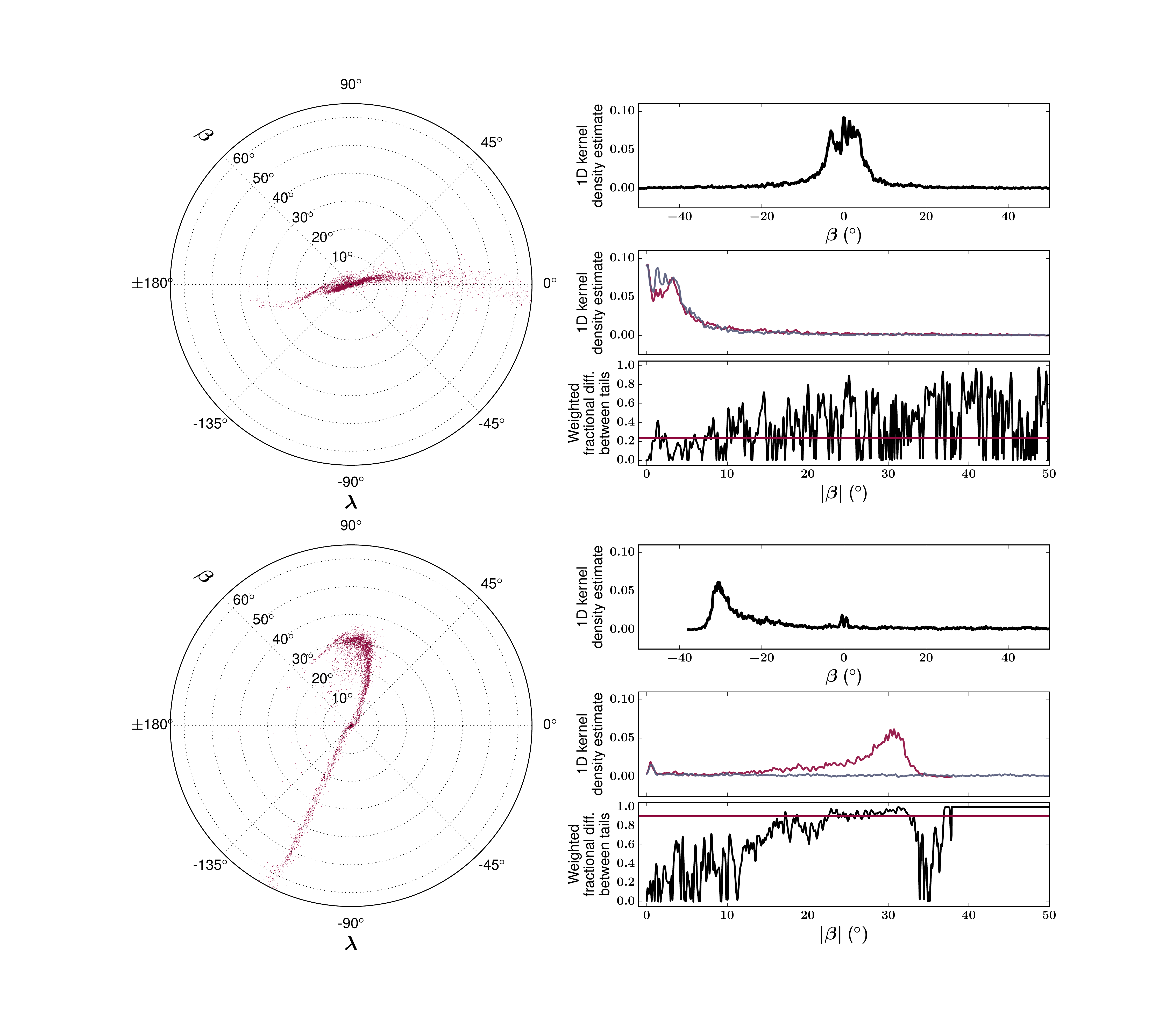}
\caption{Symmetrical (top) and asymmetrical (bottom) streams evolved in the VL2 potential. For each stream, we show the on-sky surface density (left) and the linear density profile (upper right). In each middle right panel, we plot the linear density profile folded about the globular cluster progenitor position at $\beta = 0$; the leading tail is shown in blue, and the trailing tail in red. In the lower right panel, we plot the fractional difference, $\beta$ interval by $\beta$ interval, between the leading and trailing tails (black) and the mean fractional difference weighted by the number of particles per interval (red). For the symmetrical stream, this average fractional difference is relatively small ($< 0.5$), and for the asymmetrical stream, it is relatively large ($> 0.5$). The asymmetry in this asymmetrical stream is mainly due to the density peak in the trailing tail at $\beta \simeq 30^{\circ}$, which is itself due to the trailing tail being at apocenter and exaggerated by the projection onto the $\beta$ coordinate.}
\label{fig:methods_symmetry_combined}
\end{center}
\end{figure*}

To judge stream symmetry, we again use the linear stream density profiles measured in~\ref{subsec:densityprof}. We reflect the density profiles about the globular cluster position at $\beta = 0$ and calculate the mean fractional difference between the leading and trailing tails at the center of each $0.01^{\circ}$ interval in $|\beta|$, weighted by the number of particles in the interval. This weighted average fractional difference ranges from 0 in the case of a perfectly symmetrical stream to 1 in the case of a one-tailed stream. In intervals where one of the tails has no particles, we set the fractional difference equal to 1; in intervals where both tails contain no particles, we set it equal to 0. Examples of symmetrical and asymmetrical streams are shown in fig.~\ref{fig:methods_symmetry_combined}. 

\subsubsection{Path regularity}\label{subsubsec:straightness}
\begin{figure*}
\begin{center}
\includegraphics[width=\textwidth]{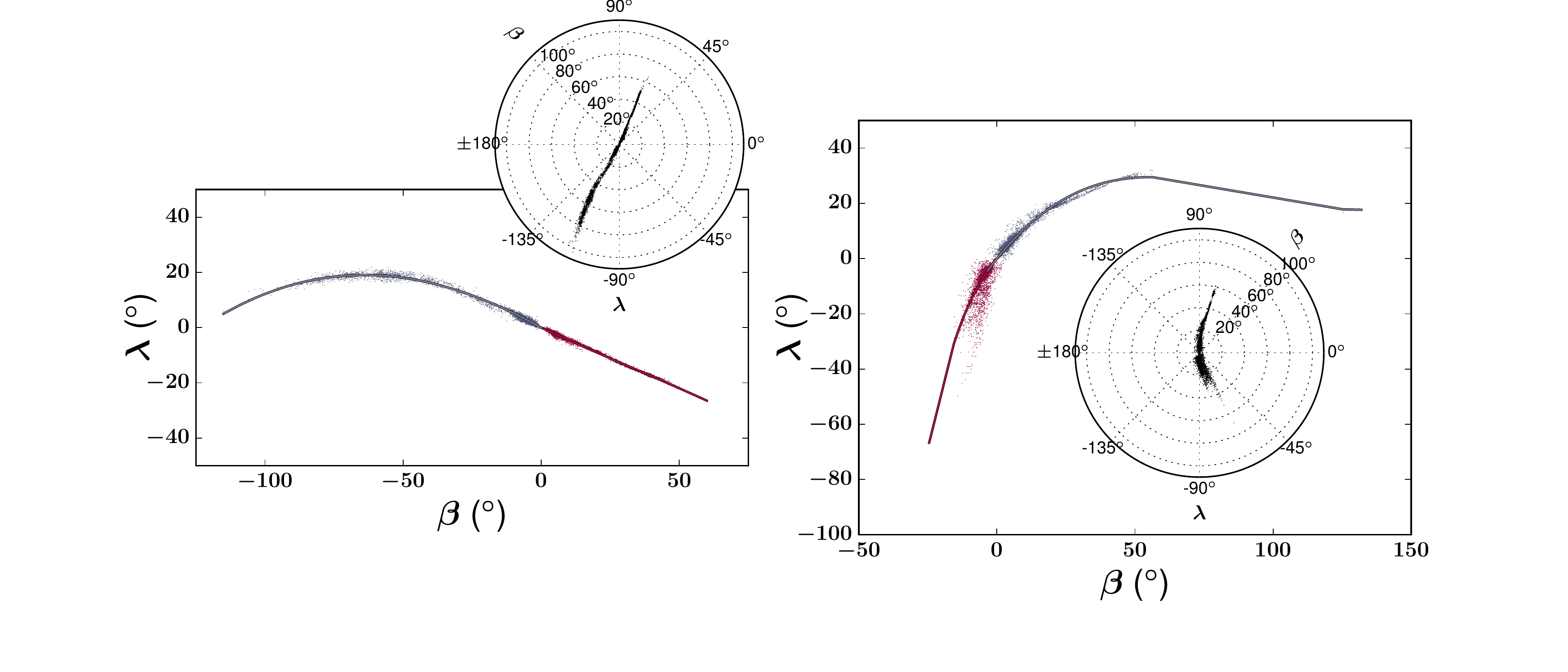}
\caption{Two streams evolved in the VL2 potential. Leading trail particles are plotted in blue, and trailing tail particles are plotted in red. The third-order polynomial fits to the leading and trailing tails are likewise plotted as blue and red lines. The left-hand stream is well-fit by this third-order polynomial model, while the right-hand stream is not. Insets: The same streams shown in polar projection.}
\label{fig:methods_straightness}
\end{center}
\end{figure*}

Finally, we quantify the simplicity of each stream's on-sky path by fitting third-order polynomials in $(\lambda,\beta)$ to the leading and trailing tails. We weight these polynomial fits by the one-dimensional kernel density estimate evaluated at each particle position; the effect of this weighting is that particles in denser regions have more of an influence on the fit than particles in sparser regions. We further constrain the polynomial fits by requiring that they intersect the globular cluster particle position, i.e. setting the constant term in the third-order polynomial equal to $0$. Examples of good and poor fits are shown in fig.~\ref{fig:methods_straightness}. 

To quantify the deviation of the tails from their respective third-order polynomial fits, we calculate the residual sum of squares (RSS) of the two tail fits and add them together to produce an overall RSS. The larger this RSS, the greater the stream's deviation from a simple polynomial path on-sky.

\subsection{Background star contamination} \label{subsec:noise}
We also consider what degree of stream disruption would be obvious above a threshold of background noise. If we could only detect the very densest parts of the stream, would ``disrupted" and ``intact" streams still quantitatively differ?

We introduce noise in the form of a background star count drawn from a Gaussian distribution and added to each $\beta$ interval in the linear stream density profile. The mean and standard deviation of this distribution are chosen to achieve a target signal-to-noise ratio of stream stars to background stars, where we define:

\begin{equation}
\mathrm{SNR} = \frac{N_{stream}}{\sqrt{N_{background}}}
\end{equation}

Here, $N_{stream}$ is the maximum number of stream stars observed in any $0.01^{\circ}\ \beta$ interval, excluding intervals within $0.5^{\circ}$ of the central globular cluster. We adopt a signal-to-noise ratio of 10, which is comparable to the SNR of the highest peak in the density profile of the Palomar 5 stream as observed by \cite{ibata16}, and solve for $N_{background}$. $N_{background}$ is then used as the mean of the Gaussian distribution from which the background star count per interval is drawn, and $\sqrt{N_{background}}$ as its standard deviation. 

We reevaluate the stream length, thinness, number of significant gaps, symmetry, and path regularity on the parts of the stream which would be ``detectable" above this background star count, where the ``detectable" regions are those with clustered high-density intervals. 

More specifically, a part of the stream is considered ``detectable" if it contains more than 3 linear density profile intervals within $1^{\circ}$ of each other where the number of stars in the interval is more than $3\sigma$ above the mean background star count (i.e., exceeds $N_{background} + 3\sqrt{N_{background}}$). 

We note that we add the background star count to the linear density profile and not the two-dimensional on-sky map of particle positions. Two of our measures (thinness and path regularity) require two-dimensional stream coordinates to compute, so for these measures, we take  the ``detectable" length of the stream to be the range of $\beta$ between the outermost detectable intervals.

\section{Results} \label{sec:results}

Figure~\ref{fig:results} presents the degree of stream disruption as parametrized by each of the five measures described in section~\ref{sec:analysis}. Each panel represents one measure; the rows represent the VL2, triaxial smooth, and spherical smooth potentials, and the columns represent the cases with and without background noise. The streams are binned by orbital location in the potential, parametrized by the maximal orbital apocentric radius and minimal orbital pericentric radius of the globular cluster progenitor particle over the 6 Gyr of stream evolution. We do not sort streams by present-day orbital phase; each bin contains streams at a range of orbital phases. Each hexagonal bin contains at least 3 streams, but the bins close to the galactic center contain, on average, many more streams than the bins far away (see fig~\ref{fig:spatialbins}). Note that the 1280 streams were evolved to match the same final conditions (namely, position and velocity of the progenitor particle) in each of the three potentials, but that the differences between the potentials caused their orbits to differ slightly, meaning that in some cases they end up in different bins. 

Here, we discuss the trends across the three potentials for each of the five measures. Later, in section~\ref{subsec:discussion_pal5}, we specifically consider the level of disruption among streams on orbits similar to that of the Palomar 5 stream: we define ``Palomar 5-like streams" as those with pericentric and apocentric radii that are both within 2.5 kpc of Palomar 5's $r_{peri} = 7.4$ kpc and $r_{apo} = 19$ kpc, from \cite{kupper15}. There are at least 40 such streams in each of the three potentials.

\subsection{Stream geometry} \label{subsec:results_geometry}

\begin{figure*}
\begin{center}
\includegraphics[width=0.75\textwidth]{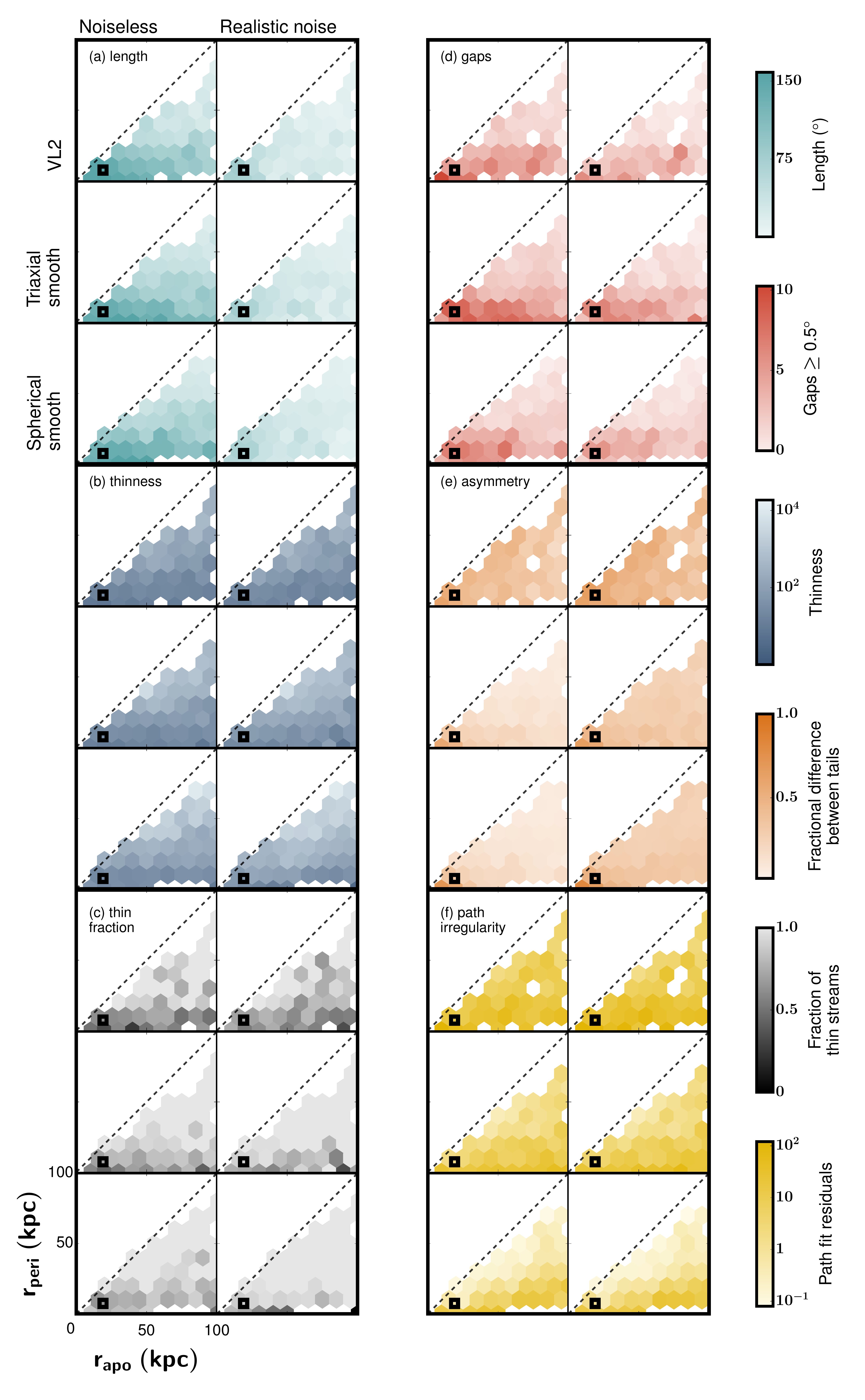}
\caption{The degree of stream disruption in each of the three potentials (VL2, smooth triaxial, smooth spherical) in each of the background noise conditions (SNR = $\infty$, SNR = 10). Each panel of six subplots represents a different measure of stream disruption. In all panels, more saturated color corresponds to greater disruption. The orbital location of the Palomar 5 stream is marked with a black box in each subplot; the outer boundary of this box encloses the orbital properties deemed ``similar" to those of the Pal 5 stream.}
\label{fig:results}
\end{center}
\end{figure*}

First, we consider trends in stream length and thinness on-sky. In fig.~\ref{fig:results}, panel (a), we plot the measurable length of the stream in each potential under each noise condition. In the case of perfect observations with no background noise, the measurable length is the full length of the stream, and in the case of realistic background, it is the length which is detectable at a $3\sigma$ level above the noise. In each subplot, we see that stream length is strongly negatively correlated with apocentric radius and strongly positively correlated with eccentricity. Both trends occur because streams which have completed more orbits (at the same energy) are longer; while stream length also depends on orbital phase and depth within the global halo potential well, these effects are secondary to number of completed orbits. Streams at small apocentric radii complete more orbits in 6 Gyr of evolution and grow to longer lengths. Meanwhile, at a constant $r_{apo}$, streams on more eccentric orbits (i.e., closer to the $r_{peri} = 0$ axis) have lower energy and a shorter orbital period than those on more circular orbits (closer to the $r_{apo} = r_{peri}$ line) and thus complete more orbits. There are no significant differences between the VL2 and smooth potential streams with respect to length. 

As background stars are added, the length of stream visible at $3\sigma$ above the mean background star count decreases; the ``detectable" region of the stream is restricted to a small, high-density region around the globular cluster progenitor particle. Far from the galactic center ($r_{apo} > 30$ kpc), the streams evolved in the spherical potential are the longest, and the streams evolved in the VL2 potential are the shortest. There is no clear trend near the galactic center.

In fig.~\ref{fig:results}, panel (b), we plot the ratio of the variance in on-sky position explained by the first principal component to that explained by the second principal component. A larger ratio indicates that the stream is thinner on-sky, and a smaller ratio indicates that the steam is bloblike and may have been dynamically heated. In each potential, we find that streams on more circular orbits are thinner than streams on more eccentric orbits, regardless of apocentric radius. In general, streams evolved in the spherical smooth potential are the thinnest, and streams in the VL2 potential are, encouragingly, the most dispersed in two dimensions on-sky. These results agree with earlier demonstrations that streams evolved in oblate or triaxial potentials are broader than streams evolved in spherical potentials (see e.g. \citealt{ibata01,mayer02,helmi04,johnston05,erkal16b}).

This trend, in which streams become less thin from the spherical to the triaxial to the VL2 potential, is most pronounced for streams on circular orbits at intermediate $r_{apo}$ (30-80 kpc). Adding background noise to the streams makes their measured variance ratio larger, indicating that the limited stream regions which are detectable above the background noise are thinner than the streams overall. 

In addition to evaluating the explained variance ratio of the PCA components for every stream in our sample, which yields a continuous measure of thinness, we impose a thinness cutoff and evaluate the \textit{fraction} of ``thin" streams in each potential. In fig.~\ref{fig:results}, panel (c), we plot the fraction of streams in each orbital bin in each potential with PCA explained variance ratio $> 10$. In all three potentials, streams on eccentric orbits are less likely to be thin than streams on circular orbits, in agreement with fig.~\ref{fig:results}, panel (b). There is no trend between the fraction of thin streams per bin and apocentric radius. Across $r_{apo}$ and $r_{peri}$, the fraction of thin streams per bin is greatest in the spherical potential and smallest in the VL2 potential. We discuss further this fraction of ``surviving" streams, i.e. streams which are still observed to be thin in the present day, in ~\ref{subsec:obs}.

In the sections below, we restrict our analyses (gap-finding, symmetry analysis, and polynomial fitting to measure path regularity) to streams which exceed the thinness cutoff, as these types of analyses would only be applied to thin streams observed in the Galaxy. This is the reason for the ``data gaps" in the VL2 panels of fig.~\ref{fig:results}, panels (d) through (f).

\subsection{Gaps} \label{subsec:results_gaps}

Next, we search for gaps in density along the stream, the theoretically predicted signatures of close subhalo flybys. In fig.~\ref{fig:results}, panel (d), we plot the total number of significant gaps detected for each stream across $0.5^{\circ}, 1.0^{\circ}, 2.0^{\circ},$ and $5.0^{\circ}$ scales. We choose to discard the gaps detected at $0.1^{\circ}$ and $0.2^{\circ}$ scales because adding background stars to each $0.01^{\circ}$ interval leads to $\sim 10$ spurious gap detections per stream at $0.1^{\circ}$ scales and $\sim 1$ spurious gap detection per stream at $0.2^{\circ}$ scales in the SNR = 10 case, as determined by running the gap-finding algorithm on a stream composed of pure noise at this level. Fig.~\ref{fig:noise_only} displays the results of one gap-finding trial on a pure-noise stream. We note that the number of spurious detections at small scales is comparable to or greater than the typical number of real gaps at larger scales. Disregarding gaps at small scales is further motivated by \cite{erkal16}, who predict that subhalo-induced gaps will be larger than $1^{\circ}$ in scale.

After restricting our analysis to gaps $\geq 0.5^{\circ}$, we find that streams on more eccentric orbits in all three potentials have more significant gaps than streams on more circular orbits. We find that the apocentric compression and pericentric stretching experienced by streams on eccentric orbits does not create over- or underdensities along streams, but that apocentric compression can obscure them, causing a stream observed at apocenter to appear smoother than the same stream observed at pericenter; this effect may actually cause us to overestimate the smoothness of streams on eccentric orbits, because we are more likely to observe any given stream at its apocenter than its pericenter.

Streams at smaller apocentric radii have more gaps than streams at larger apocentric radii. Counterintuitively, streams in the triaxial smooth potential have the most gaps overall, while streams in the spherical smooth potential and the VL2 potential have comparable numbers of significant gaps. 

We find that increasing the level of background noise results in fewer significant gap detections; this is the result of large gaps in the no-background case being smoothed out by the noise, as well as a shorter length of the stream being ``detectable" above increasing levels of noise. We note in particular that more than twice as many significant gaps are detected for Palomar 5-like streams in the \textit{smooth} potentials ($5.0^{+6.2}_{-3.0}$ gaps each) than in the lumpy VL2 potential ($2.0^{+3.0}_{-2.0}$ gaps each) (see~\ref{subsec:discussion_pal5} for further discussion). The above trends all persist when we restrict our analysis even further, to gaps $> 1^{\circ}$, to compare to the predictions of \cite{erkal16}, although fewer gaps are detected at these largest scales overall.

\begin{figure*}
\begin{center}
\includegraphics[width=\textwidth]{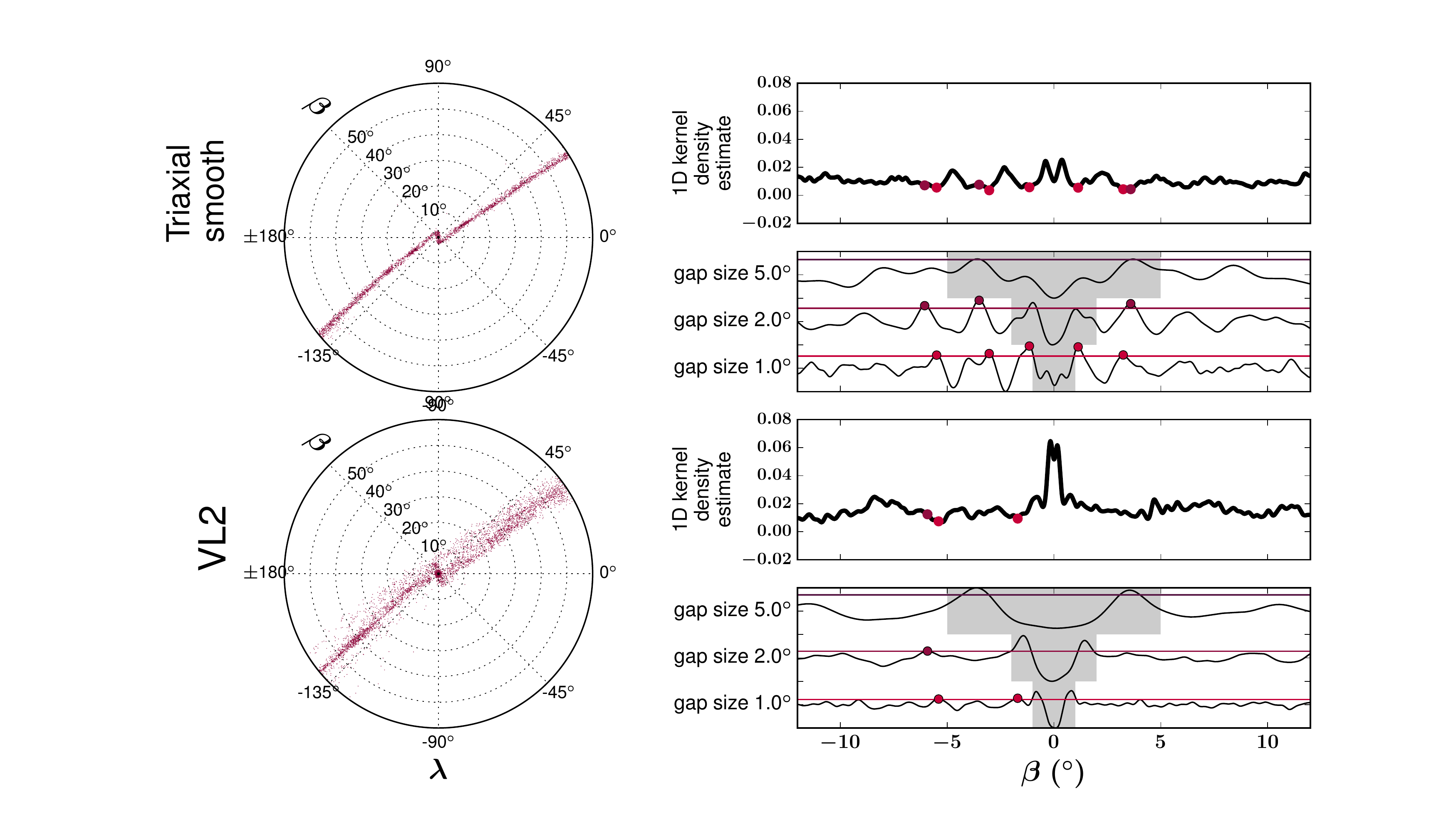}
\caption{A representative comparison of the gaps detected in a VL2 stream (bottom) and the triaxial smooth potential stream evolved to match it (top). For each stream, we show the linear density profile and mark the gaps detected at $1.0^{\circ}, 2.0^{\circ}$, and $5.0^{\circ}$ scales. Below, we show the convolution with the $w_2$ filter and the significance threshold for each gap scale. The stream evolved in the triaxial smooth potential has several regularly spaced, symmetric gaps which are caused by the epicyclic motions of the stream particles. The VL2 stream, which is messier and more dynamically heated, lacks these neat gaps. As a result, there are fewer gaps detected overall in the VL2 stream than its smooth potential counterpart.}
\label{fig:qualitativegaps}
\end{center}
\end{figure*}

\begin{figure*}
\begin{center}
\includegraphics[width=\textwidth]{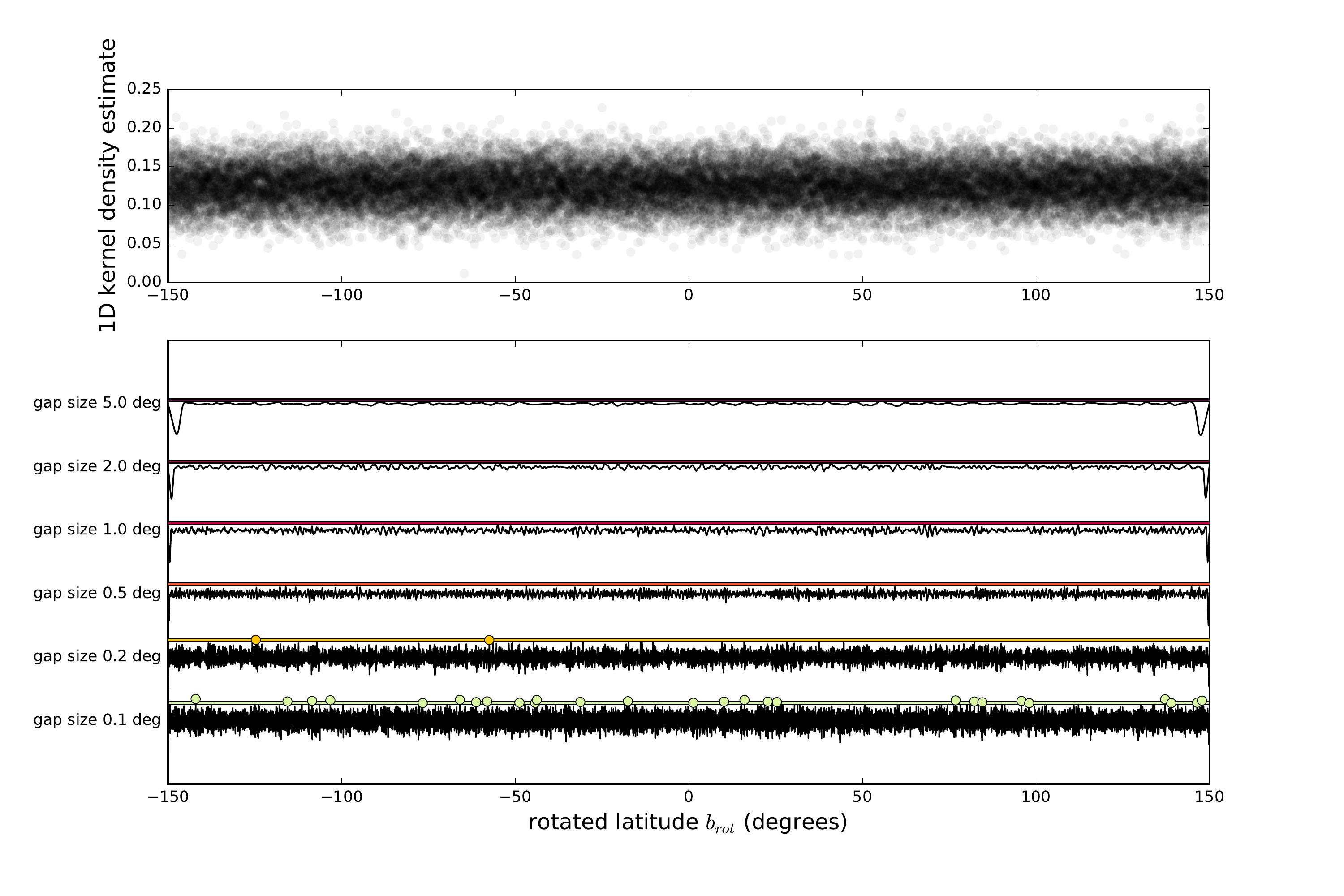}
\caption{A simulated ``noise-only" stream, where background stars are added to a stream of uniform density at a level consistent with SNR = 10. In a typical realization of such a noise-only stream, $\sim 10$ significant gaps are detected at the $0.1^{\circ}$ scale, and $\sim 1$ significant gaps are detected at the $0.2^{\circ}$ scale.}
\label{fig:noise_only}
\end{center}
\end{figure*}

We further investigate the counterintuitive result that the VL2 streams have fewer gaps, on average, than their smooth potential counterparts by comparing the gaps detected in individual VL2 streams to those detected in the corresponding triaxial smooth potential streams. Fig.~\ref{fig:qualitativegaps} shows a representative stream comparison. In general, fewer gaps are detected in the VL2 streams because neat epicyclic over- and underdensities, which are readily detected by the gap-finding procedure described in~\ref{subsubsec:gaps}, do not survive in the VL2 potential.

We emphasize that the detected gaps are qualitatively different across the three potentials, and individual gaps still tell us much about the physical processes affecting stream evolution. In fig.~\ref{fig:methods_gaps_gappy}, for example, we plot a stream evolved in the spherical smooth potential with several detected gaps, but these symmetrical, evenly spaced gaps are plainly the result of epicyclic density variations \citep{kupper08}. Streams evolved in the VL2 potential, in general, lack neat epicyclic gaps and have instead ``messier," less symmetrical, non-evenly spaced gaps. Previous work has demonstrated that subhalo-induced gaps bear detailed dynamical signatures of the responsible subhalo perturbation, so this messiness is information-rich \citep{erkal15properties,sanders16}). The point is that convolving the stream density profile with a gap-shaped filter cannot distinguish between differently-shaped gaps, and simply counting the number of significant gap detections resulting from this procedure does not capture the important qualitative differences between the three stream populations. The overall gap count is a poor measure of the lumpiness of the halo potential.

\subsection{Symmetry} \label{subsec:results_symmetry}

We also investigate the symmetry of the leading and trailing tails of our simulated streams. In fig.~\ref{fig:results}, panel (e), we plot the mean fractional difference in density, $0.01^{\circ} |\beta|$ interval by interval, between the leading and trailing tails of the stream, weighted by the number of stars per interval. We find that streams at low apocentric radius are less symmetrical than streams at high apocentric radius, and that for streams evolved in the two smooth potentials, streams on eccentric orbits are less symmetrical than streams on circular orbits. This eccentricity trend is likely due to the density variations arising from stream compression and stretching at apocenter and pericenter, respectively; see fig.~\ref{fig:methods_gaps_gappy} for an example of an asymmetric stream from the spherical smooth potential with an obvious asymmetry resulting from apocentric compression of the leading tail. Interestingly, stream symmetry does not correlate with orbital eccentricity for the VL2 streams. 

In the case of no background stars, VL2 streams at all $r_{apo}$ and $r_{peri}$ are considerably less symmetrical than their smooth potential counterparts. We emphasize that this conclusion holds in spite of the averaging over the present-day orbital phases of the streams in each bin. The differences between the VL2 streams and the smooth potential streams are, once again, starkest for streams on circular orbits at $r_{apo} > 30$ kpc.

Adding background noise erases the formerly obvious differences between the VL2 and smooth potentials. Streams in all three potentials are observed to be less symmetrical in the SNR = 10 case than in the case of no background stars, but the difference is more dramatic for smooth potential streams at $r_{apo} > 30$ kpc. 

\subsection{Path regularity} \label{subsec:results_straightness}

Finally, we investigate the ``regularity" or simplicity of the on-sky paths of the leading and trailing tails of our simulated streams. In fig.~\ref{fig:results}, panel (f), we plot, on a logarithmic scale, the residual sum of squares (RSS) between the stream particle positions and a third-order polynomial fit. The fit is weighted by the density of stream particles, such that particles in dense regions of the stream have more influence over the fit than particles in sparse regions. We find that, in every potential, orbital eccentricity is strongly positively correlated with RSS, i.e. that orbital eccentricity is negatively correlated with goodness-of-fit of the third-order polynomial. There does not appear to be a trend between RSS and apocentric radius. 

In general, the spherical smooth potential streams are best fit by the third-order polynomial model, then the triaxial smooth potential streams, and then the VL2 streams. This is true across all $r_{apo}$ and eccentricity, although the differences between the potentials are once again starkest for streams on circular orbits and $r_{apo} \sim 30-80$ kpc. For such streams, the median RSS of the third-order polynomial fit differs by two orders of magnitude between the spherical smooth potential and the VL2 potential. Encouragingly, adding background noise does not affect these results, likely because the stream particles which become invisible above the background noise are those in the sparser regions of the stream, which, because of the weighting, have only minimal influence over the polynomial fit even in the case of no background stars.

\section{Discussion} \label{sec:discussion} 

\subsection{Comparisons to the observed Palomar 5 stream} \label{subsec:discussion_pal5}

The Palomar 5 stream is perhaps the most scrutinized globular cluster stream in the Galaxy, and Pal 5 served as the model for our choice of globular cluster progenitor parameters. Here, we consider the subset of Cauda streams on ``Palomar 5-like" orbits and ask: how do our simulated Palomar 5-like streams, ``observed" against a realistic level of background noise, compare to the observed Palomar 5 stream? A strong similarity to one of the three simulated stream populations would not be conclusive evidence for or against lumpiness in the Milky Way halo potential, but it would at least indicate correspondence between our simulated streams and reality.

The observed Palomar 5 stream is at least $22^{\circ}$ long on-sky \citep{gd06_pal5,ibata16}. According to fig. 7a of \cite{ibata16}, the extent of the Palomar 5 stream, projected into the plane of the sky, is roughly 11.7 kpc; the authors find that the characteristic width of the stream is only 58 pc, which means the length-to-width ratio of the stream is $\sim 200$. As we indicated in the introduction to this work, there is disagreement in the literature with regard to the number of significant gaps present in the Palomar 5 stream, but the two published analyses of the most recent photometric observations of this stream which focused on identifying individual gaps claim 0 and 2 gaps, respectively \citep{ibata16,erkal16arxiv}, with the latter gaps measuring $2^{\circ}$ and $9^{\circ}$. The fractional difference in binned star count between the leading and trailing tails, weighted by the number of stars per $0.1^{\circ}$ bin, is 0.69 based on figure 7b of \cite{ibata16}. This high degree of asymmetry in the Palomar 5 stream has been noted before: \cite{erkal16arxiv} conclude that the asymmetry near the stream progenitor must be due to a gravitational perturbation in the stream's history, possibly from the Galactic bar, and \cite{pearson17} demonstrate that the dramatic difference in the leading and trailing tail lengths could also be due to bar-induced perturbations. Finally, \cite{ibata16} find that the on-sky paths of both the leading and trailing tails of the Palomar 5 stream are well-fit by third-order polynomials; it is difficult to estimate the goodness-of-fit of these models, but their residuals show no large-scale trends.

We define ``Palomar 5-like" streams in our simulations as those with $r_{peri}$ and $r_{apo}$ that are both within 2.5 kpc of Palomar 5's $r_{peri} = 7.4$ kpc and $r_{apo} = 19$ kpc, respectively, from \cite{kupper15}. There are at least 40 such streams in each of the three potentials. In each panel of fig.~\ref{fig:results}, a black box marks the location of the Palomar 5-like stream subset; the outer border of the box encloses the full subset, and the inner border marks the position of Pal 5 itself. 

There are no significant differences between Pal 5-like streams in the three potentials in terms of length, regardless of the level of background noise. At SNR=10, comparable to the peak SNR of the recent \cite{ibata16} observations of the real Pal 5, our simulated Pal 5-like streams are $\sim 60^{\circ}$ long, on average. Without background noise, Pal 5-like streams are equally thin in all three potentials (explained variance ratio $\simeq 20$), but when background noise is added, the distinguishing prospects improve somewhat: Pal 5-like streams are still comparably thin in the VL2 and triaxial smooth potentials (explained variance ratio $\simeq 50$), but more than twice as thin in the spherical smooth potential (explained variance ratio $ > 100$). Streams appear thinner when background noise is added because, typically, the ends of the tails are broader and more dispersed than the central regions. When noise is added, the stream appears shorter overall because the dispersed outer parts are lost in the background noise, but the visible section close to the progenitor is thinner than the stream as a whole, and its small width more than compensates for the lost length.

Correspondingly, we find that the fraction of Palomar 5-like streams which are thin is between 50 and 60\% in all three potentials as long as the streams are observed without background noise. When a realistic level of background noise is added, more streams are ``thin" by the cutoff overall, but the differences between potentials are more pronounced: 73\% of Palomar 5-like streams in the VL2 potential are ``thin," while 93\% of streams in the triaxial smooth potential and 89\% of streams in the spherical smooth potential are thin. These differences are of only modest ($\sim 1\sigma$) significance given the number of Palomar 5-like streams in each potential ($\sim 40$).

In the case of no background noise, Palomar 5-like streams have, typically, $4.5^{+4.2}_{-3.2}$ significant gaps each in the VL2 potential and $8.5^{+7.7}_{-4.7}$ significant gaps each in the smooth potentials. (We note again that these counts represent the total number of significant gaps over $0.5^{\circ}$, $1.0^{\circ}$, $2.0^{\circ}$, and $5.0^{\circ}$ scales; here, we report the median count $\pm 1 \sigma$.) The differences between the potentials remain consistent when noise is added, even as the large-scale gaps are washed out: in the lumpy VL2 potential, streams have $2.0^{+3.0}_{-2.0}$ gaps each, and in the smooth potentials, they have $5.0^{+6.2}_{-3.0}$ gaps each. Overall, therefore, the Palomar 5-like streams evolved in the \textit{smooth} potentials have roughly twice as many gaps as those evolved in the VL2 potential at roughly $1\sigma$ significance.

Palomar 5-like streams follow the general trends of our simulated sample in terms of symmetry; they are considerably less symmetrical in the VL2 potential (weighted fractional difference between leading and trailing tails $ = 0.53^{+0.09}_{-0.19}$) than the smooth potentials (weighted fractional difference between tails $ = 0.36^{+0.20}_{-0.17}$) when no background noise is added. Streams in all three potentials become somewhat less symmetrical when noise is added, but the differences between the potentials diminish: the weighted fractional difference between the tails is $0.61^{+0.12}_{-0.22}$ in the VL2 potential, and $0.45^{+0.20}_{-0.13}$ in the smooth potentials. Regardless of the level of noise, Pal 5 streams have the most regular paths in the spherical smooth potential (median RSS $\sim 10$) and the least regular paths in the VL2 potential (median RSS $\sim 100$).


Very broadly, the observed Palomar 5 stream is $\sim 40^{\circ}$ shorter than the average of our simulated Palomar 5-like streams in all three potentials, although the leading tail of the observed Palomar 5 stream overhangs the edge of the SDSS footprint, so there is reason to believe that it may be longer than present observations suggest. 
The observed Palomar 5 stream is most consistent with the streams evolved in the spherical potential in terms of thinness, although it is broadly consistent with all three potentials. The observed Palomar 5 stream is most consistent with the VL2 streams in terms of the number of gaps and dramatic leading-trailing tail asymmetry. The general goodness of the \cite{ibata16} third-order polynomial fit to the on-sky path of the Palomar 5 stream is most consistent with the smooth spherical halo potential.




\subsection{Measures of stream disruption}
\label{subsec:discussion_measures}

Of our five measures of stream disruption---length, thinness, gaps, symmetry, and path regularity---we find that thinness, symmetry, and path regularity are most informative with regard to distinguishing between the three potentials. Simple gap counting yields the counterintuitive result that the VL2 streams are smoother than streams evolved in the spherical and triaxial smooth potentials. However, more detailed analysis of the gaps in individual streams reveals that many of the gaps detected in the smooth potential streams are epicyclic in nature, and that easily detectable epicyclic gaps do not survive in the VL2 potential (see fig.~\ref{fig:qualitativegaps}). This may also offer an explanation as to why large-scale gaps are disproportionately obscured by background noise in the VL2 potential as compared to the smooth potentials---because VL2 gaps are less like the symmetrical, regular $w_2$ filter than smooth-potential gaps, they generate a weaker signal in the convolution which is less likely to stand out above background noise.

Altogether, our results suggest that traditional measures of stream disruption, especially gap counting, may be insufficient to characterize disrupted vs. intact streams, and that we should consider applying more creative measures to observed Milky Way streams. Beyond this work, a very promising approach in this regard is that of \cite{bovy16}, who examine the power spectra of fluctuations in density and ``track," or on-sky path. This approach exploits the same information in density profile and on-sky path that we have used here, but it additionally overcomes several limitations of gap-finding by wavelet analysis: it is not limited to searching for gaps at discrete scales, and it can capture the complex signatures of multiple overlapping subhalo interactions, which do not yield clean, gap-filter-shaped density perturbations. \cite{bovy16} find that the density and track power spectra of perturbed streams are sensitive to the overall mass spectrum of perturbing subhalos, with large-scale power resulting mainly from large subhalo flybys and vice versa.

\cite{bovy16} also undertake an investigation of the density profile of the Palomar 5 stream as observed by \cite{ibata16}. In their figure 30, they present the power spectrum of fluctuations in this density profile; they approximate the noise level along the stream as SNR = 2 by taking the median signal-to-noise ratio of the density profile. (We, in comparison, adopt the maximum signal-to-noise ratio, 10, when generating our ``noisy" streams, but similarly find that small-scale gaps, which translate to small-scale power, are completely obscured by the noise in the density profile.) At large gap scales ($\sim 5^{\circ}$ and above), \cite{bovy16} find that the power spectrum of the density profile is consistent with that predicted by a CDM-like subhalo population. Applying a similar power spectrum analysis to our simulated streams would provide an alternative picture of their density structure, and could yield other distinctions between the Cauda and smooth potential streams, but this is beyond the scope of the present study.

\subsection{Ranking stream disruption measures with machine learning}
To demonstrate the relative potential-distinguishing power of our five measures of stream disruption, we train a machine learning algorithm to predict the parent potential of an input stream based on its measured disruption, then rank the importance or weighting of the five measures to the algorithm's ultimate prediction. More specifically, we summarize each stream as an array of ten numerical ``features:"

\begin{enumerate}
\item[1  ] Length: the stream length, in degrees
\item[2  ] Thinness: the PCA explained variance ratio
\item[3-8  ] Gaps: the number of detected gaps at $0.1$, $0.2$, $0.5$, $1.0$, $2.0$, and $5.0^{\circ}$ scales
\item[9 ] Symmetry: The weighted fractional difference between the leading and trailing tails
\item[10  ] Path regularity: The RSS of a third-order polynomial fit to the on-sky paths of the leading and trailing tails
\end{enumerate}

The machine learning algorithm attempts to ``learn" a relationship between these input features and the stream's parent potential. We train the algorithm on a randomly selected 70\% of the data set and evaluate its performance in classifying the remaining 30\%. 

We use a Random Forest classifier \citep{breiman01}, which averages the predictions of a large number of individual decision trees, each trained on a different random subset of the training set. We experiment with random forest hyperparameters (including the number of individual decision trees which are averaged together and varying degrees of complexity within individual trees), using a randomized parameter search and five-fold cross validation to find the parameters which yield the best classification performance.

\begin{figure*}
\begin{center}
\includegraphics[width=\textwidth]{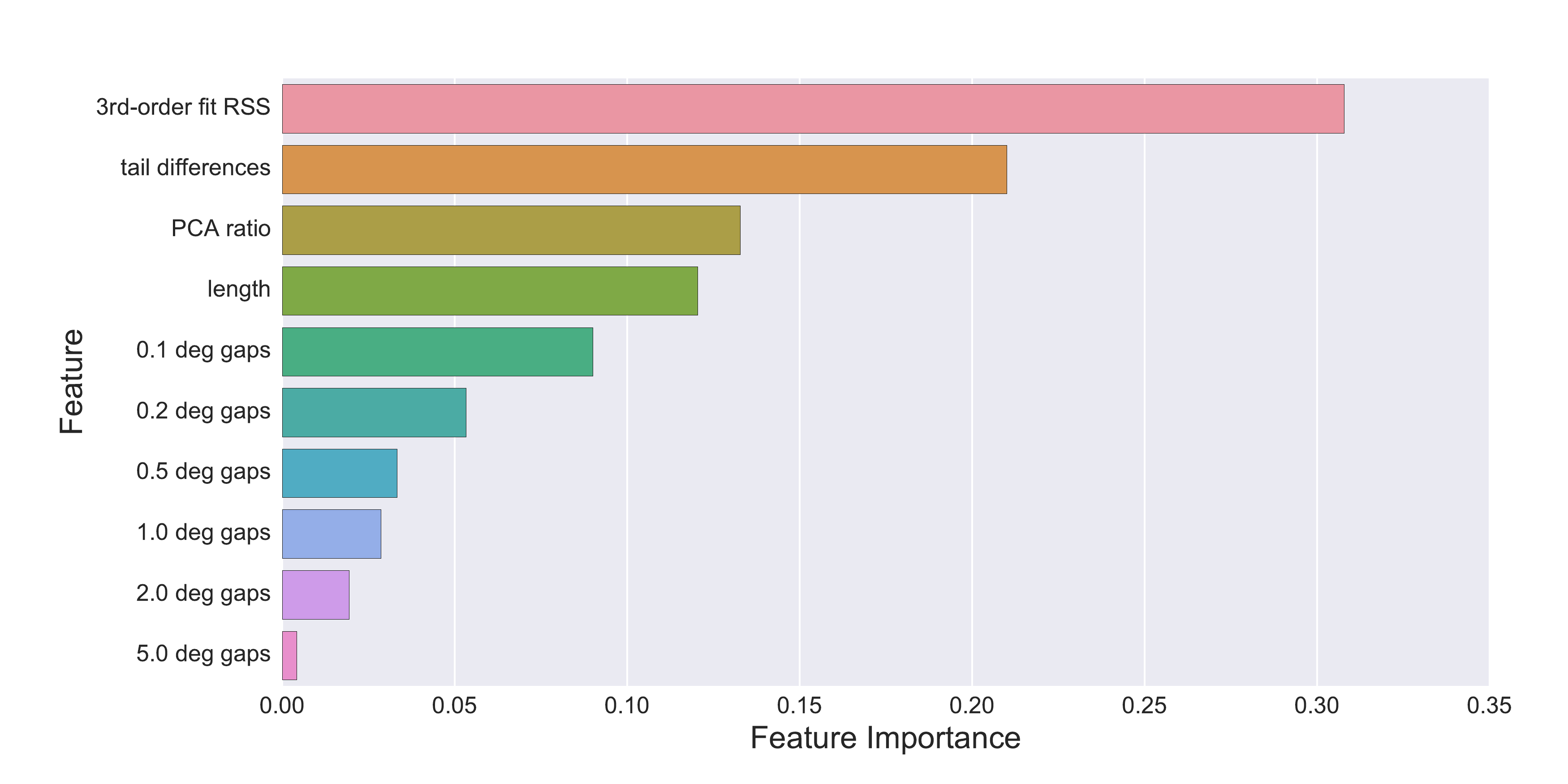}
\caption{The relative importance of input features to the random forest classifier in the ultimate classification of SNR = 10 test set streams as belonging to the VL2, smooth triaxial, or smooth spherical potential. The importances are normalized such that they sum to 1. As expected based on fig.~\ref{fig:results}, the three most important measures of stream disruption are path regularity, symmetry, and thinness.}
\label{fig:rfranking}
\end{center}
\end{figure*}

Our final model is able to consistently classify the test set streams with $\sim 65\%$ accuracy in the case of no background noise and $\sim 60\%$ accuracy in the case of SNR = 10, a roughly twofold improvement over random chance. In fig.~\ref{fig:rfranking}, we plot the relative importance of each of the ten input features to the ultimate classification made by the random forest in the SNR = 10 case. As fig.~\ref{fig:results} indicates, the three most important features are, in order, the RSS of the weighted third-order polynomial fit, the weighted fractional difference between the leading and trailing tails, and the ratio of variances explained by the PCA components. Thus, path regularity, symmetry, and thinness are the most important measures of stream disruption for distinguishing between the three potentials. Interestingly, stream length was of nearly comparable importance to the explained variance ratio from PCA, which suggests that there may be subtle structure in panel (a) of fig.~\ref{fig:results}, or that interactions of this feature with other features are important.

Finally, we group the triaxial and spherical potential streams together and test the random forest's ability to classify streams as belonging to either a ``lumpy" (VL2) or ``smooth" parent potential. In this simplified picture, the random forest is able to consistently classify the test set streams with $\sim 80\%$ accuracy in the case of no background noise and $\sim 70\%$ accuracy in the case of SNR = 10. Under both noise conditions, the feature importance ranking changes significantly from the three-potential case (see fig.~\ref{fig:rfranking})---the symmetry measure (tail differences) becomes the most important feature, followed by length, path regularity (third-order fit RSS), and thinness (PCA ratio). This change may also be understood with reference to fig.~\ref{fig:results}; the most important features in classifying streams as ``lumpy" vs. ``smooth" are those which are similar for the triaxial and smooth potential streams, but different for the VL2 streams. Symmetry (panel (e)) typifies this pattern, while thinness (panel (b)), for example, differs strongly between all three potentials.

\subsection{Future observational prospects} \label{subsec:obs}

Our results suggest that Palomar 5-like streams are relatively poorly placed to distinguish between the three potentials, but that streams on circular orbits at larger apocentric radii ($r_{apo} \sim 30-80$ kpc) bear more obvious signs of their parent potential. While it is not yet clear how many streams we might expect to exist at such large distances, if there are examples, the prospects for observing them in the near future are hopeful. The Large Synoptic Survey Telescope \citep{ivezic08}, with projected first light in 2021 \citep{LSSTsite}, will survey more than half the sky to a depth several apparent magnitudes deeper than the star-count surveys in which the tidal tails of Palomar 5 were discovered \citep{odenkirchen01}, allowing surface brightness limits equivalent to the (relatively nearby) current detections in star counts up to 100kpc away. In addition, the next generation of multi-object spectrographs are being planned with sufficient fiber density and on large enough telescopes (e.g., the Prime Focus Spectrograph on the Subaru telescope, \citealt{takada14}) that an analysis of a large number of stars in velocity space is feasible even for distant targets. Such an analysis could both remove contaminating foreground and background stars in discovered streams and add an extra dimension of constraining phase-space information.




\section{Conclusions} \label{sec:conclusions}

In this work, we consider the observable effects of dark matter substructure and global halo potential shape on the morphology of tidal streams. We present the Via Lactea Cauda streams, a population of 1280 simulated globular cluster streams evolved on a range of orbits in the Via Lactea II halo. We compare the Cauda streams to corresponding control stream populations evolved in triaxial and spherical smooth potentials fit to VL2. In particular, we evaluate the degree of morphological ``disruption" apparent in each stream population, parametrized by five measures: length, thinness, number of gaps, leading vs. trailing tail symmetry, and on-sky path regularity. 

We find that the Cauda streams, evolved in the VL2 potential, are by some measures far more ``disrupted" than their smooth potential counterparts: their on-sky paths are significantly more complex (i.e., poorly fit by simple polynomials), their leading and trailing tails are dramatically less symmetrical, and they are less thin on-sky. Surprisingly, by the traditional measure of stream disruption, the Cauda streams are more intact than corresponding smooth-potential streams: they have \textit{fewer} individual significant gaps, as found by convolution with a gap-shaped filter. We emphasize that this result holds only for simple gap-counting; it is likely that the gaps identified in smooth potential streams are epicyclic in nature, while pristine epicyclic gaps and overdensities may not survive in the complicated VL2 potential. Furthermore, the clarity and strength of epicyclic overdensities depends on the nature and orbit of the disrupting globular cluster progenitor (see e.g. \citealt{lane12}). Detailed analysis of the shape and regularity of individual gaps along individual streams (see e.g. \citealt{erkal15,erkal16}) is necessary to draw conclusions about the origin of gaps, and indeed, if a gap is subhalo-induced, its precise shape is rich in information about the perturbing subhalo (see e.g. \citealt{erkal15properties,sanders16}).

We therefore find that stream thinness, symmetry, and path regularity are useful indicators of whether a stream's parent potential is lumpy or smooth, but gap counting is less conclusive and should be treated with caution. This does not mean, however, that statistical treatments of density variations along streams cannot indicate the character of the parent potential (see e.g. \citealt{bovy16}). Stream length, in our analysis, is not affected by the character of the halo potential and indicates only the number of orbits the stream has completed.

We furthermore find that these measures are most useful for distinguishing between smooth and lumpy potentials when evaluated on globular cluster streams on circular orbits at intermediate apocentric radii ($\sim 30-80$ kpc). Streams orbiting nearer to the galactic center, like the Palomar 5 stream, are less dramatically affected by their parent halo potential and therefore less promising for distinguishing between a lumpy and smooth Milky Way halo. This holds even in our simulations, in which there is no galactic disk to destroy subhalos at small galactocentric radii.

Finally, we emphasize the critical importance of accounting for background star contamination---the level of background noise can make more of a difference than the shape of the potential in a stream's ultimate observed morphology. This is especially true for the detectable length of a stream and its observed degree of leading-trailing tail symmetry. Stream thinness and path regularity are less affected by background noise, so we should apply these measures to observed Milky Way streams if possible, although measuring the thinness of real streams will be observationally difficult, because it requires observations of streams beyond a narrow angular range about their central path. 

Overall, while we encourage caution in interpreting existing stream observations---namely, careful choice of methods of analysis and accounting for background noise---there are encouraging prospects, both photometrically and spectroscopically, that streams at larger distances might be discovered and mapped in high dimensions in the near future. Upcoming observations of both known and new streams will significantly improve upon existing observations and enhance our ability to distinguish potentials. 

\section*{Acknowledgements}

We thank the referee for detailed and thoughtful comments. ES thanks Z. Penoyre for useful discussions and graphic design insight. KVJ and ES were supported by NSF grants AST-1312196 and AST-1614743 and NASA grant NNX15AK78G. AHWK would like to acknowledge support through DFG Research Fellowship KU 3109/1-1 and from NASA through Hubble Fellowship grant HST-HF-51323.01-A awarded by the Space Telescope Science Institute, which is operated by the Association of Universities for Research in Astronomy, Inc., for NASA, under contract NAS 5-26555.



\bibliographystyle{mnras}
\bibliography{streamsbib}

\bsp	
\label{lastpage}
\end{document}